\documentclass[aps,prl,twocolumn,groupedaddress]{revtex4-1}
\usepackage{graphicx}
\usepackage{amsmath}
\usepackage{amssymb}
\usepackage{braket}
\usepackage{bm}
\newtheorem{thm}{Theorem}
\newtheorem{remark}{Prescription}
\begin{document}

\title{A Hessian Geometric Structure of Chemical Thermodynamic Systems with Stoichiometric Constraints}
%\author{Yuki Sughiyam$\mathrm{a}^{1}$, Dimitri Loutchk$\mathrm{o}^{1}$, Atsushi Kamimur$\mathrm{a}^{1}$ and Tetsuya J. Kobayash$\mathrm{i}^{1}$}
\author{Yuki Sughiyama, Dimitri Loutchko, Atsushi Kamimura and Tetsuya J. Kobayashi}
%\affiliation{${}^{1}\mathrm{I}$nstitute of Industrial Science, The University of Tokyo, 4-6-1, Komaba, Meguro-ku, Tokyo 153-8505 Japan}
\affiliation{Institute of Industrial Science, The University of Tokyo, 4-6-1, Komaba, Meguro-ku, Tokyo 153-8505 Japan}
%\affiliation{${}^{2}\mathrm{P}$RESTO, Japan Science and Technology Agency (JST), 4-1-8, Honcho, Kawaguchi, Saitama 332-0012 Japan}
\date{\today}
\begin{abstract}
We establish a Hessian geometric structure in chemical thermodynamics which describes chemical reaction networks (CRNs) with equilibrium states. 
In our setup, the ideal gas assumption and mass action kinetics are not required.  
The existence and uniqueness condition of the equilibrium state is derived by using the Legendre duality inherent to the Hessian structure. 
The entropy production during a relaxation to the equilibrium state can be evaluated by the Bregman divergence. 
Furthermore, the equilibrium state is characterized by four distinct minimization problems of the divergence, which are obtained from the generalized Pythagorean theorem originating in the dual flatness. 
For the ideal gas case, we confirm that our existence and uniqueness condition implies Birch's theorem, 
and that the entropy production represented by the divergence coincides with the generalized Kullback-Leibler divergence. 
In addition, under mass action kinetics, our general framework reproduces the local detailed balance condition.
\end{abstract}
%\pacs{87.23.Kg, 87.10.Mn 05.70.Ln, 05.40.-a}
\maketitle
\section{I. Introduction}
Chemical thermodynamics is a solid physical basis for treating systems consisting of chemical reactions \cite{01,02,03,a1}. 
In recent years, it has found new applications in biophysics and systems biology \cite{b1,b2,b3,b4,b5,b6,b7}, and also been actively extended by incorporating new techniques from stochastic thermodynamics \cite{04,05,06,07,08,09}, chemical reaction network theory \cite{m4,m5,m6} and information theory \cite{10}.
However, from a general physics perspective, chemical thermodynamics is a chimera of pure thermodynamic and kinetic aspects.

Historically, the conventional equilibrium chemical thermodynamics was established by the seminal papers by Gibbs \cite{a2}, 
in which the chemical equilibrium state is variationally and globally characterized as the state to minimize the free energy. 
In the same period, the chemical kinetic theory was also being developed in parallel with Gibbs' equilibrium chemical thermodynamics.

By combining the law of mass action by Guldberg and Waage in 1864 \cite{c3} with Boltzmann's characterization of the equilibrium state by detailed balancing, 
Wegscheider clarified the condition which the rate constants of chemical reactions must satisfy to have equilibrium states \cite{c4}. 
The characterization of an equilibrium state by the detailed balancing of the reaction fluxes is kinetic and local but consistent with the global free-energy characterization of the equilibrium state under the ideal-gas or dilute-solution assumption \cite{c5}.

Since then, theories of chemical reaction systems and their thermodynamics have been developed, mainly based on the detailed balancing characterization of equilibrium states and mass action systems. 
For example, in the 1970s, the chemical reaction network (CRN) theory emerged \cite{c2}. Here, Horn and Jackson formalized the complex-balanced state of a mass action system, which extended the uniqueness and stability of the equilibrium state \cite{m1}. 
In relation to the stability of the equilibrium state, it was found that the Gibbs free energy difference is identical to the generalized Kullback-Leibler divergence (also known as the pseudo-Helmholtz potential) 
and behaves as a Lyapunov function of mass action systems \cite{a3,a4,a5,m1}. 
This result could be regarded as a chemical version of Boltzmann’s H theorem and was used for proving the convergence of a mass action system to the unique equilibrium state characterized by detailed balancing \cite{a6,a7}. 
Also, around 1970, Hill and Schnakenberg extended the theory to stochastic linear reaction systems \cite{03,a8,m2}.

The applicability of chemical thermodynamics has recently been extended further in various ways. 
The generalized mass action kinetics was proposed in the field of applied mathematics as a broader class of kinetics in which the properties of the equilibrium state can be conserved \cite{m3,m4,m5,m6,a9}. 
The authors of Refs. \cite{06,07,08,09,c1} established a theory for open CRNs and derived the conditions under which an open system has an equilibrium state . 

However, most of recent developments are not purely thermodynamic, because they are based on the characterization of the equilibrium state by detailed balancing, assuming a specific kinetics analogous to the mass action. 
As a result, it is unclear whether the results are obtained just by mathematical extensions that crucially depend on the specific kinetics of the mass action and its variants, 
or whether they are truly consistent with the general framework of  thermodynamics. 
If they are consistent, the same results must be derived from a purely thermodynamic argument in the line of Gibbs without assuming kinetics and detailed balancing. 
Once the consistencies are confirmed, one could apply the previously-established results to a much broader class of non-ideal and non-mass action chemical systems, because thermodynamics can describe the properties of systems independent of the details of their kinetics. However, there have been few attempts to establish the link \cite{a9}. In this work, we reveal the link in chemical thermodynamics of general CRNs.

Before outlining our main results, we recall the general framework of thermodynamics, which should be formulated as follows. 
The space of extensive variables is endowed with a concave function, called entropy \cite{01,a1}. According to the second law of thermodynamics, a system should evolve with time such that the total entropy function increases under any given constraints imposed on the extensive variables. 
If the constraints are trivial, the system converges to the maximum of the total entropy function, namely, the equilibrium state. 
Furthermore, there is a well-established procedure to evaluate the dissipation during a relaxation to the equilibrium state.

However, the constraints can have non-trivial impacts when such a general theory is applied to CRNs, especially to those with complex stoichiometry of chemical reactions. 
In a chemical thermodynamic system, the extensive variables include the numbers of molecules constituting the system.
These numbers cannot change independently as they are algebraically constrained by the stoichiometry.
It is these constraints which provide significant geometric structure into the problem, and also 
yield several properties obtained from detailed balancing and mass action kinetics. 
%Nevertheless, we expect that the general thermodynamic framework can derive the results obtained by detailed balancing and the mass action kinetics without assuming any of them.

{\bf [Outline of Main Results]}
In this paper, we develop a thermodynamic theory for chemical reaction systems with complex constraints. With this theory, we obtain the following four main results.

%\begin{itemize}
{\bf Theorem 1. Necessary and Sufficient Condition for Existence of Equilibrium States for Open CRNs:}
This is a generalization of the Wegscheider condition and the equilibrium condition of open CRNs obtained recently in Refs. \cite{06,07}.

{\bf Theorem 2. Uniqueness Condition of the Equilibrium State:}
This is a generalization of the uniqueness condition of the equilibrium state obtained so far under the assumption of mass action kinetics.

{\bf Theorem 3. The Difference of Total Entropy between the Equilibrium State and Any State is Evaluated by Bregman Divergence:}
This is a generalization of the fact that the free energy difference is identical to the generalized Kullback-Leibler divergence. In particular, we clarify that a convex function characterizing the Bregman divergence corresponds to the thermodynamic potential of the system.

{\bf Theorem 4. Four Variational Characterizations of the Equilibrium State:}
One of them is the generalization of the variational characterization of the equilibrium state as the minimizer of the free energy (the generalized Kullback-Leibler divergence). The other three are newly obtained as a result of our Hessian geometric formulation.
%\end{itemize}

We emphasize again that these results are derived purely thermodynamically, without using any specific kinetics such as the mass action laws and the local characterization of equilibrium states such as detailed balancing. 
In particular, we derive these generalizations by identifying and employing the Hessian geometric structure \cite{c6} in constrained thermodynamic systems. 
The Hessian geometry of thermodynamic systems plays an essential role, when the constraints between the variables become non-trivial and complex. 

This paper is organized as follows.
We devote Sec. II to review the conventional thermodynamics of CRNs in light of the entropy maximization problem. Also, we recapitulate how different kinds of thermodynamic potentials are linked to each other.  
In Sec. III, we derive the existence and uniqueness condition for the equilibrium state. To this purpose, we introduce two spaces which are connected to each other by Legendre duality. This pair of the spaces and their duality is the basis of the Hessian geometric structure.
The equilibrium state is then uniquely determined by the intersection of two submanifolds (see Theorem 2). 
Sec. IV reformulates the second law from the geometric point of view. We show that the dissipation can be evaluated by the Bregman divergence (see Theorem 3). 
Furthermore, we obtain four distinct characterizations of the equilibrium state (see Theorem 4). 
In Sec. V, we demonstrate our geometric structure in ideal gas cases to rederive the previously known results: 
our theorems for the existence and uniqueness condition reduce to Birch's theorem (see Theorem 5), and the entropy production can be represented by the generalized Kullback-Leibler divergence. 
In addition, by assuming the law of mass action, we reproduce the local detailed balance condition. 
Finally, we summarize our results with further discussions in Sec. VI.  

While we derive all the results without the assumptions of mass action systems and detailed balancing, the assumptions are familiar to researchers working on stochastic thermodynamics and CRNs. 
In our accompanying paper \cite{KobaAccompaning}, we reproduce our results for the special case starting from mass action systems and detailed balancing.

\section{II. Conventional thermodynamics for chemical reaction systems}

In this section, we recall conventional thermodynamics for chemical reaction systems. 
Readers who are familiar with the topic can skip to Eqs. (\ref{enttotXi}), (\ref{heatphi}) and (\ref{VPXi}). 

Consider a thermodynamic chemical reaction system surrounded by a reservoir.
We assume that the system is always in a local equilibrium state, i.e., a well-mixed state, and therefore we can completely describe it by extensive variables $\left(\Omega,E,N,X\right)$. 
Here, $\Omega$ and $E$ represent the volume and the internal energy; $N=\left\{N^{m}\right\}$ denotes a vector, each component of which is the number of the corresponding open chemical.  The open chemicals can diffuse across the boundary between the system and the reservoir. 
By contrast, $X=\left\{X^{i}\right\}$ is the numbers of chemicals confined in the system; the indices $m$ and $i$ run from $m=1$ to $\mathcal{N}_N$ and from $i=1$ to $\mathcal{N}_X$, respectively, where $\mathcal{N}_N$ and $\mathcal{N}_X$ represent the numbers of species of the open and confined chemicals. 
Since we only discuss isochoric cases (i.e., $\Omega=\mathrm{const.}$) for theoretical simplicity \cite{n0}, we employ the density variables $\left(\epsilon,n,x\right)=\left(E/\Omega,N/\Omega,X/\Omega\right)$. 
In thermodynamics, a concave, smooth and homogeneous function $\Sigma$, which is called the entropy, is defined on $\left(\Omega,E,N,X\right)$. 
Owing to the homogeneity of the entropy function, without loss of generality, we can write it as 
\begin{equation}
\Sigma\left[\Omega,E,N,X\right]=\Omega\sigma\left[\epsilon,n,x\right],\label{fe}
\end{equation}
where $\sigma\left[\epsilon,n,x\right]$ represents the entropy density. 
In this work, we additionally assume that $\sigma\left[\epsilon,n,x\right]$ is strictly concave, which implies a situation without phase transitions from the physical point of view. 
The reservoir is characterized by intensive variables $(\tilde{T},\tilde{\mu})$, where $\tilde{T}$ is temperature and $\tilde{\mu}=\left\{\tilde{\mu}_{m}\right\}$ are chemical potentials corresponding to the open chemicals; 
also we denote the corresponding extensive variables by $(\tilde{E},\tilde{N})$. 
We denote the entropy function for the reservoir by $\tilde{\Sigma}_{\tilde{T},\tilde{\mu}}[\tilde{E},\tilde{N}]$, and therefore the total entropy can be expressed by 
\begin{equation}
\Sigma^{\mathrm{t}\mathrm{o}\mathrm{t}}\left[\epsilon,n,x|\tilde{E},\tilde{N}\right]=\Omega\sigma\left[\epsilon,n,x\right]+\tilde{\Sigma}_{\tilde{T},\tilde{\mu}}\left[\tilde{E},\tilde{N}\right],\label{enttot}
\end{equation} 
where we use the additivity of the entropy. 

\begin{figure}[h]
\begin{center}
\includegraphics[width=0.5\textwidth]{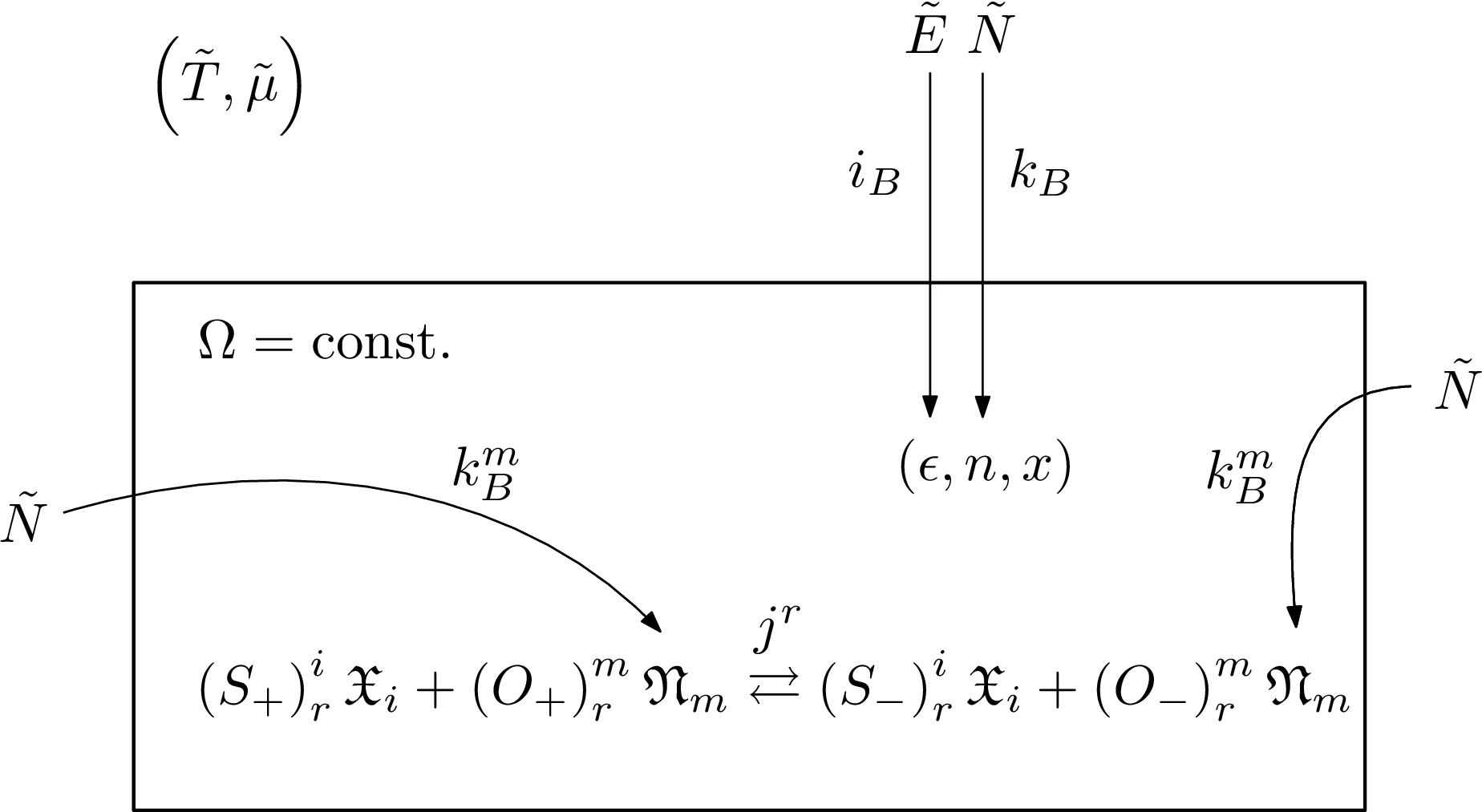}
\caption{Diagrammatic explanation of an open CRN. 
The chemical reactions occur with the reaction flux densities $j(t) = \{ j^r(t) \}$, the $r$th reaction of which is represented as the chemical equation in the figure. 
Here, $\mathfrak{X}=\left\{\mathfrak{X}_{i}\right\}$ are the labels of the confined chemicals, and $\mathfrak{N}=\left\{\mathfrak{N}_{m}\right\}$ are the ones of the open chemicals, which can move across the boundary with the diffusion flux densities $k_B(t) = \{ k_B^m(t) \}$.
Also, $\left(S_{+}\right)_{r}^{i}$ and $\left(O_{+}\right)_{r}^{m}$ denote stoichiometric coefficients of the reactants in the $r$th reaction, 
whereas $\left(S_{-}\right)_{r}^{i}$ and $\left(O_{-}\right)_{r}^{m}$ are the ones of the products. Thus, the stoichiometric matrices are given as $S_{r}^{i}=\left(S_{-}\right)_{r}^{i}-\left(S_{+}\right)_{r}^{i}, O_{r}^{m}=\left(O_{-}\right)_{r}^{m}-\left(O_{+}\right)_{r}^{m}$. }
\label{fig1}
\end{center}
\end{figure}

Next, we define a dynamics as 
\begin{eqnarray}
\displaystyle \nonumber&&\frac{d\epsilon}{dt}=i_{B}\left(t\right),\mbox{  }\frac{dn^{m}}{dt}=O_{r}^{m}j^{r}\left(t\right)+k_{B}^{m}\left(t\right),\mbox{  }\frac{dx^{i}}{dt}=S_{r}^{i}j^{r}\left(t\right),\\
&&\displaystyle \frac{d\tilde{E}}{dt}=-\Omega i_{B}\left(t\right),\mbox{  }\frac{d\tilde{N}^{m}}{dt}=-\Omega k_{B}^{m}\left(t\right),\label{Dynamics}
\end{eqnarray}
where $i_{B}\left(t\right),\ j\left(t\right)=\left\{j^{r}\left(t\right)\right\}$ and $k_{B}\left(t\right)=\left\{k_{B}^{m}\left(t\right)\right\}$ represent the energy, the chemical reaction and the chemical diffusion flux densities, respectively; 
also, $S=\left\{S_{r}^{i}\right\}$ and $O=\left\{O_{r}^{m}\right\}$ denote stoichiometric matrices for the confined and open chemicals (see FIG. \ref{fig1} and also Eqs. (\ref{CheEq}) and (\ref{SOmat})). 
The index $r$ runs from $r=1$ to $\mathcal{N}_{R}$, where $\mathcal{N}_{R}$ is the number of reactions. 
In this paper, we employ Einstein's summation convention for notational simplicity. 

Since, in most cases, the time scale of reactions is much slower than the others (that is, $i_{B}\left(t\right),k_{B}\left(t\right)\gg j\left(t\right)$), 
we can analyze the dynamics, Eq. (\ref{Dynamics}), by separating it into the fast and slow scales. 
By employing a scaling: $\tau:=\alpha t,\ i\left(\tau\right):=\alpha^{-1}i_{B}\left(t\right),\ k\left(\tau\right):=\alpha^{-1}k_{B}\left(t\right),\ \alpha\rightarrow\infty$, we obtain the fast scale effective dynamics as 
\begin{eqnarray}
\displaystyle \nonumber&&\frac{d\epsilon}{d\tau}=i\left(\tau\right),\mbox{  }\frac{dn^{m}}{d\tau}=k^{m}\left(\tau\right),\\
&&\displaystyle \frac{d\tilde{E}}{d\tau}=-\Omega i\left(\tau\right),\mbox{  }\frac{d\tilde{N}^{m}}{d\tau}=-\Omega k^{m}\left(\tau\right),\label{FDynamics}
\end{eqnarray}
where we use the fact that the diverging bare flux densities, $i_{B}$ and $k_{B}$, converge to the finite effective flux densities, $i$ and $k$, in the scaling limit. 
The formal solution of Eq. (\ref{FDynamics}) with an initial condition $(\epsilon_{0},n_{0},\tilde{E}_{0},\tilde{N}_{0})$ can be represented as  
\begin{eqnarray}
\nonumber&&\epsilon\left(\tau\right)=\epsilon_{0}+\iota\left(\tau\right),\mbox{  }n^{m}\left(\tau\right)=n_{0}^{m}+\kappa^{m}\left(\tau\right),\\
&&\tilde{E}\left(\tau\right)=\tilde{E}_{0}-\Omega\iota\left(\tau\right),\mbox{  }\tilde{N}^{m}\left(\tau\right)=\tilde{N}_{0}^{m}-\Omega\kappa^{m}\left(\tau\right),\label{Fevo}
\end{eqnarray}
where $\iota\left(\tau\right)$ and $\kappa\left(\tau\right)$ are the integrals of $i\left(\tau\right)$ and $k\left(\tau\right)$ with the initial condition $\iota\left(0\right)=\kappa\left(0\right)=0$:
\begin{equation}
    \iota\left(\tau\right) = \int_{0}^{\tau} i\left(\tau'\right) d\tau', \mbox{  } 
     \kappa^m\left(\tau\right) = \int_{0}^{\tau} k^m\left(\tau'\right) d\tau'.
\end{equation}
Here, we note that the densities of confined chemicals, $x$, can be regarded as a constant in the fast dynamics. 
By substituting the solution, Eq. (\ref{Fevo}), into Eq. (\ref{enttot}), we have the time evolution of total entropy as 
\begin{eqnarray}
\nonumber\Sigma^{\mathrm{t}\mathrm{o}\mathrm{t}}&&\left(\iota\left(\tau\right),\kappa\left(\tau\right),x\right)=\Omega\sigma\left[\epsilon_{0}+\iota\left(\tau\right),n_{0}+\kappa\left(\tau\right),x\right]\\
&&-\displaystyle \frac{1}{\tilde{T}}\Omega\iota\left(\tau\right)+\frac{\tilde{\mu}_{m}}{\tilde{T}}\Omega\kappa^{m}\left(\tau\right)+\mathrm{const}.,\label{Fenttot}
\end{eqnarray}
where we use properties of the reservoir, $\Omega\iota\left(t\right)\ll\tilde{E}_{0},\ \Omega\kappa\left(t\right)\ll\tilde{N}_{0}$, and the Taylor expansion for $\tilde{\Sigma}_{\tilde{T},\tilde{\mu}}$; we also employ the thermodynamic relations: $\partial\tilde{\Sigma}_{\tilde{T},\tilde{\mu}}/\partial\tilde{E}=1/\tilde{T}$ and $\partial\tilde{\Sigma}_{\tilde{T},\tilde{\mu}}/\partial\tilde{N}^{m}=-\tilde{\mu}_{m}/\tilde{T}$. 
Although the constant term is explicitly given as $\tilde{\Sigma}_{\tilde{T},\tilde{\mu}}[\tilde{E}_{0},\tilde{N}_{0}]$, we abbreviate it to ``$\mathrm{const}.$", because it never affects the theoretical framework. 

To introduce thermodynamics into our dynamics, we briefly summarize its significant statements. 
According to the first law, a heat dissipation $\mathcal{Q}_{0\rightarrow\tau}$ from the system to the reservoir during a time interval $\left[0,\tau\right]$ is given by the entropy increment in the reservoir: 
\begin{eqnarray}
\nonumber&&\mathcal{Q}_{0\rightarrow\tau}:=-\Omega\iota\left(\tau\right)+\Omega\tilde{\mu}_{m}\kappa^{m}\left(\tau\right)\\
\nonumber&&=\tilde{T}\left\{\tilde{\Sigma}_{\tilde{T},\tilde{\mu}}\left(\tau\right)-\tilde{\Sigma}_{\tilde{T},\tilde{\mu}}\left(0\right)\right\}\\
&&=\tilde{T}\left\{\Sigma^{\mathrm{t}\mathrm{o}\mathrm{t}}\left(\tau\right)-\Sigma^{\mathrm{t}\mathrm{o}\mathrm{t}}\left(0\right)\right\}-\tilde{T}\Omega\left\{\sigma\left(\tau\right)-\sigma\left(0\right)\right\},\label{Fheat}
\end{eqnarray}
where $\Omega\iota\left(\tau\right)$ represents the internal energy gain of the system and $-\Omega\tilde{\mu}_{m}\kappa^{m}\left(\tau\right)$ is the work done by the system through the injection of chemicals into the reservoir. 

The second law states that, for spontaneous changes, the flux density functions, $i\left(\tau\right)$ and $k\left(\tau\right)$, must be chosen such that $\Sigma^{\mathrm{t}\mathrm{o}\mathrm{t}}\left(\tau\right)$ becomes an increasing function with respect to time $\tau$ \cite{n1}. 
In other words, the system climbs up the landscape defined by the concave function $\Sigma^{\mathrm{t}\mathrm{o}\mathrm{t}}\left(\iota,\kappa,x\right)$ with respect to $\iota$ and $\kappa$ in the time evolution, and finally converges to the maximum if it exists. 
If we write $\left(\epsilon\left(\tau\right),n\left(\tau\right)\right)\rightarrow\left(\epsilon_{\mathrm{Q}\mathrm{E}\mathrm{Q}},n_{\mathrm{Q}\mathrm{E}\mathrm{Q}}\right)$ for $\tau \rightarrow \infty$, candidates of the converged state $\left(\epsilon_{\mathrm{Q}\mathrm{E}\mathrm{Q}},n_{\mathrm{Q}\mathrm{E}\mathrm{Q}}\right)$ can be evaluated by a variational form: 
\begin{eqnarray}
\displaystyle \nonumber&&\left(\iota_{\mathrm{Q}\mathrm{E}\mathrm{Q}},\kappa_{\mathrm{Q}\mathrm{E}\mathrm{Q}}\right)\in\arg\max_{\iota,\kappa}\Sigma^{\mathrm{t}\mathrm{o}\mathrm{t}}\left(\iota,\kappa,x\right)\\
&&=\displaystyle \arg\max_{\iota,\kappa}\left\{\sigma\left[\epsilon_{0}+\iota,n_{0}+\kappa,x\right]-\frac{1}{\tilde{T}}\iota+\frac{\tilde{\mu}_{m}}{\tilde{T}}\kappa^{m}\right\},
\end{eqnarray}
and $\left(\epsilon_{\mathrm{Q}\mathrm{E}\mathrm{Q}},n_{\mathrm{Q}\mathrm{E}\mathrm{Q}}\right)=\left(\epsilon_{0}+\iota_{\mathrm{Q}\mathrm{E}\mathrm{Q}},n_{0}+\kappa_{\mathrm{Q}\mathrm{E}\mathrm{Q}}\right)$. 
In thermodynamics, the states maximizing the total entropy are called equilibrium states; therefore, $\left(\epsilon_{\mathrm{Q}\mathrm{E}\mathrm{Q}},n_{\mathrm{Q}\mathrm{E}\mathrm{Q}}\right)$ is an equilibrium state in the fast dynamics. 
However, we call it a quasi-equilibrium state, because we will treat the slow dynamics later. 
By using the argument shift, $\epsilon=\epsilon_{0}+\iota, n=n_{0}+\kappa$, we can rewrite the variational form as
\begin{equation}
\displaystyle \left(\epsilon_{\mathrm{Q}\mathrm{E}\mathrm{Q}},n_{\mathrm{Q}\mathrm{E}\mathrm{Q}}\right)\in\arg\max_{\epsilon,n}\left\{\sigma\left[\epsilon,n,x\right]-\frac{1}{\tilde{T}}\epsilon+\frac{\tilde{\mu}_{m}}{\tilde{T}}n^{m}\right\},\label{VPEN}
\end{equation}
and we directly obtain candidates of the quasi-equilibrium state. 

Since, in thermodynamics, the function maximized in Eq. (\ref{VPEN}) is bounded above  \cite{n2}, the quasi-equilibrium state always exists.
Furthermore, since we have assumed the strict concavity for $\sigma$ in this work, we can conclude that the quasi-equilibrium state is uniquely determined by Eq. (\ref{VPEN}); 
and, for an arbitrary initial condition $\left(\epsilon_{0},n_{0}\right)$, the system converges to the unique quasi-equilibrium state $\left(\epsilon_{\mathrm{Q}\mathrm{E}\mathrm{Q}},n_{\mathrm{Q}\mathrm{E}\mathrm{Q}}\right)$. 
The above conclusion, which is the existence and uniqueness of the quasi-equilibrium state, originates from the simplicity of the fast dynamics, Eq. (\ref{FDynamics}). 
In other words, the maximization is easily conducted, because $\epsilon$ and $n$ can be varied independently.
As shown later, the conclusion no longer holds for the slow reaction dynamics, because of complex stoichiometric constraints. 
Also, the total entropy at the quasi-equilibrium state can be represented as 
\begin{equation}
\displaystyle \Sigma_{\mathrm{Q}\mathrm{E}\mathrm{Q}}^{\mathrm{t}\mathrm{o}\mathrm{t}}=\Omega\max_{\epsilon,n}\left\{\sigma\left[\epsilon,n,x\right]-\frac{1}{\tilde{T}}\epsilon+\frac{\tilde{\mu}_{m}}{\tilde{T}}n^{m}\right\}+\mathrm{const}.\label{entQEQ}
\end{equation}

Employing the above results for the fast dynamics, we analyze the slow dynamics, which is the chemical reaction dynamics. 
Owing to the variational form, Eq. (\ref{VPEN}), the time evolutions of the densities of the internal energy $\epsilon\left(t\right)$ and of the open chemicals $n\left(t\right)$ in the slow dynamics are already solved. By using the time evolution of the confined chemicals $x\left(t\right)$, we have 
\begin{equation}
\epsilon\left(t\right)=\epsilon_{\mathrm{Q}\mathrm{E}\mathrm{Q}}\left(\tilde{T},\tilde{\mu};x\left(t\right)\right),\mbox{  }n\left(t\right)=n_{\mathrm{Q}\mathrm{E}\mathrm{Q}}\left(\tilde{T},\tilde{\mu};x\left(t\right)\right).\label{Sen}
\end{equation} 
Substituting these equations into Eq. (\ref{Dynamics}), we obtain the effective slow dynamics as
\begin{eqnarray}
\displaystyle \nonumber&&\frac{dx^{i}}{dt}=S_{r}^{i}j^{r}\left(t\right),\mbox{  }\frac{d\tilde{E}}{dt}=-\Omega\frac{d\epsilon_{\mathrm{Q}\mathrm{E}\mathrm{Q}}\left(\tilde{T},\tilde{\mu};x\left(t\right)\right)}{dt},\\
&&\displaystyle \frac{d\tilde{N}^{m}}{dt}=\Omega\left\{O_{r}^{m}j^{r}\left(t\right)-\frac{dn_{\mathrm{Q}\mathrm{E}\mathrm{Q}}^{m}\left(\tilde{T},\tilde{\mu};x\left(t\right)\right)}{dt}\right\}.\label{SDynamics}
\end{eqnarray}
The formal solution of Eq. (\ref{SDynamics}) with the initial condition $x_{0}$ is represented as 
\begin{eqnarray}
\nonumber x^{i}\left(t\right)&=&x_{0}^{i}+S_{r}^{i}\xi^{r}\left(t\right),\\
\nonumber\tilde{E}\left(t\right)&=&\tilde{E}\left(0\right)-\Omega\epsilon_{\mathrm{Q}\mathrm{E}\mathrm{Q}}\left(\tilde{T},\tilde{\mu};x\left(t\right)\right),\\
\nonumber\tilde{N}^{m}\left(t\right)&=&\tilde{N}^{m}\left(0\right)+\Omega\left\{O_{r}^{m}\xi^{r}-n_{\mathrm{Q}\mathrm{E}\mathrm{Q}}\left(\tilde{T},\tilde{\mu};x\left(t\right)\right)\right\},\\\label{Sevo}
\end{eqnarray}
where $\xi\left(t\right)=\left\{\xi^{r}\left(t\right)\right\}$ is the integral of $j\left(t\right)$ with the initial condition $\xi\left(0\right)=0$. The vector $\xi\left(t\right)$ is the density of the extent of reaction. 
Also, the initial conditions of the reservoir for the slow dynamics, $\tilde{E}\left(0\right)$ and $\tilde{N}\left(0\right)$, can be calculated from the fast dynamics as
\begin{eqnarray}
\nonumber\tilde{E}\left(0\right)&=&\tilde{E}_{0}-\Omega\left(\epsilon_{\mathrm{Q}\mathrm{E}\mathrm{Q}}\left(\tilde{T},\tilde{\mu};x_0\right)-\epsilon_{0}\right),\\
\tilde{N}^{m}\left(0\right)&=&\tilde{N}_{0}^{m}-\Omega\left(n_{\mathrm{Q}\mathrm{E}\mathrm{Q}}^{m}\left(\tilde{T},\tilde{\mu};x_0\right)-n_{0}^{m}\right).
\end{eqnarray}
The substitution of Eqs. (\ref{Sen}) and (\ref{Sevo}) into Eq. (\ref{enttot}) leads to the time evolution of the total entropy in the reaction dynamics: 
\begin{eqnarray}
\nonumber&&\Sigma^{\mathrm{t}\mathrm{o}\mathrm{t}}\left(\xi\left(t\right)\right)=\Omega\sigma\left[\epsilon_{\mathrm{Q}\mathrm{E}\mathrm{Q}}\left(\xi\right),n_{\mathrm{Q}\mathrm{E}\mathrm{Q}}\left(\xi\right),x_{0}+S\xi\right]\\
\displaystyle \nonumber&&-\frac{1}{\tilde{T}}\Omega\epsilon_{\mathrm{Q}\mathrm{E}\mathrm{Q}}\left(\xi\right)-\frac{\tilde{\mu}_{m}}{\tilde{T}}\Omega\left\{O_{r}^{m}\xi^{r}-n_{\mathrm{Q}\mathrm{E}\mathrm{Q}}\left(\xi\right)\right\}+\mathrm{const}.,\\\label{Senttot}
\end{eqnarray}
where we use the Taylor expansion for $\tilde{\Sigma}_{\tilde{T},\tilde{\mu}}$ again. 
If we use the quasi-equilibrium entropy function  $\Sigma_{\mathrm{Q}\mathrm{E}\mathrm{Q}}^{\mathrm{t}\mathrm{o}\mathrm{t}}=\Sigma_{\mathrm{Q}\mathrm{E}\mathrm{Q}}^{\mathrm{t}\mathrm{o}\mathrm{t}}(\tilde{T},\tilde{\mu};x\left(t\right))$ given by Eq. (\ref{entQEQ}), we can rewrite Eq. (\ref{Senttot}) as 
\begin{equation}
\displaystyle \Sigma^{\mathrm{t}\mathrm{o}\mathrm{t}}\left(\xi\right)=\Sigma_{\mathrm{Q}\mathrm{E}\mathrm{Q}}^{\mathrm{t}\mathrm{o}\mathrm{t}}\left(\tilde{T},\tilde{\mu};x_{0}+S\xi\right)-\Omega\frac{\tilde{\mu}_{m}}{\tilde{T}}O_{r}^{m}\xi^{r}+\mathrm{const}.\label{QEQenttot}
\end{equation}

From the second law, an equilibrium state in the reaction dynamics is evaluated by a variational form: $\xi_{\mathrm{E}\mathrm{Q}}=\arg \mathrm{max}_{\xi}\Sigma^{\mathrm{t}\mathrm{o}\mathrm{t}}\left(\xi\right)$, and $x_{\mathrm{E}\mathrm{Q}}=x_{0}+S\xi_{\mathrm{E}\mathrm{Q}}$; also, the equilibrium total entropy is $\Sigma_{\mathrm{E}\mathrm{Q}}^{\mathrm{t}\mathrm{o}\mathrm{t}}=\mathrm{max}_{\xi}\Sigma^{\mathrm{t}\mathrm{o}\mathrm{t}}\left(\xi\right)$. 
Furthermore, by following the same argument as in Eq. (\ref{Fheat}), the heat dissipation of this dynamics is given by 
\begin{eqnarray}
\nonumber&&\mathcal{Q}_{t^{\prime}\rightarrow t}=\tilde{T}\left\{\Sigma^{\mathrm{t}\mathrm{o}\mathrm{t}}\left(\xi\left(t\right)\right)-\Sigma^{\mathrm{t}\mathrm{o}\mathrm{t}}\left(\xi\left(t^{\prime}\right)\right)\right\}\\
&&-\tilde{T}\Omega\left\{\sigma_{\mathrm{Q}\mathrm{E}\mathrm{Q}}\left(x\left(t\right)\right)-\sigma_{\mathrm{Q}\mathrm{E}\mathrm{Q}}\left(x\left(t^{\prime}\right)\right)\right\},\label{Sheat}
\end{eqnarray}
where $\sigma_{\mathrm{Q}\mathrm{E}\mathrm{Q}}\left(x\left(t\right)\right):=\sigma\left[\epsilon_{\mathrm{Q}\mathrm{E}\mathrm{Q}}\left(t\right),n_{\mathrm{Q}\mathrm{E}\mathrm{Q}}\left(t\right),x\left(t\right)\right]$ denotes the system entropy density at the quasi-equilibrium state with the confined chemicals $x\left(t\right)$. 

The representation of the total entropy, Eq. (\ref{QEQenttot}), may be unfamiliar to the reader, therefore we rewrite it by employing thermodynamic potentials. 
First, consider the maximization in Eq. (\ref{entQEQ}) with respect to $\epsilon$: 
\begin{equation}
\displaystyle \psi\left[\frac{1}{\tilde{T}};n,x\right]:=\max_{\epsilon}\left\{\sigma\left[\epsilon,n,x\right]-\frac{1}{\tilde{T}}\epsilon\right\},\label{TPm}
\end{equation}
which is called the Massieu potential density. 
By using this potential, the Helmholtz free-energy density is defined as 
\begin{equation}
f\left[\tilde{T};n,x\right]:=-\tilde{T}\psi\left[\frac{1}{\tilde{T}};n,x\right].\label{TPf}
\end{equation}
Finally, a variant of the Legendre transfomation \cite{n3} of $f[\tilde{T};n,x]$ leads to 
\begin{equation}
\displaystyle \varphi\left[\tilde{T},\tilde{\mu};x\right]:=\min_{n}\left\{f\left[\tilde{T};n,x\right]-\tilde{\mu}_{m}n^{m}\right\},\label{TPg}
\end{equation}
which coincides with the partial grand potential density. 
Owing to the definition of $\varphi[\tilde{T},\tilde{\mu};x]$, the quasi-equilibrium entropy function $\Sigma_{\mathrm{Q}\mathrm{E}\mathrm{Q}}^{\mathrm{t}\mathrm{o}\mathrm{t}}$ can be represented as 
\begin{equation}
\displaystyle \Sigma_{\mathrm{Q}\mathrm{E}\mathrm{Q}}^{\mathrm{t}\mathrm{o}\mathrm{t}}\left(\tilde{T},\tilde{\mu};x\right)=-\frac{\Omega}{\tilde{T}}\varphi\left[\tilde{T},\tilde{\mu};x\right]+\mathrm{const}.,
\end{equation}
and therefore Eq. (\ref{QEQenttot}) can be rewritten in a familiar form: 
\begin{equation}
\displaystyle \Sigma^{\mathrm{t}\mathrm{o}\mathrm{t}}\left(\xi\right)=-\frac{\Omega}{\tilde{T}}\left\{\varphi\left[\tilde{T},\tilde{\mu};x_{0}+S\xi\right]+\tilde{\mu}_{m}O_{r}^{m}\xi^{r}\right\}+\mathrm{const}.\label{enttotXi}
\end{equation}
Also, the differentiation of $\varphi[\tilde{T},\tilde{\mu};x]$ with respect to $\tilde{T}$ gives $-\sigma_{\mathrm{Q}\mathrm{E}\mathrm{Q}}\left(x\right)$ (See Appendix A). 
Hence, the heat dissipation, Eq. (\ref{Sheat}), can be expressed as 
\begin{eqnarray}
\nonumber&&\mathcal{Q}_{t^{\prime}\rightarrow t}=\tilde{T}\left\{\Sigma^{\mathrm{t}\mathrm{o}\mathrm{t}}\left(\xi\left(t\right)\right)-\Sigma^{\mathrm{t}\mathrm{o}\mathrm{t}}\left(\xi\left(t^{\prime}\right)\right)\right\}\\
&&+\tilde{T}\Omega\left\{\frac{\partial\varphi\left[\tilde{T},\tilde{\mu};x\left(t\right)\right]}{\partial\tilde{T}}-\frac{\partial\varphi\left[\tilde{T},\tilde{\mu};x\left(t^{\prime}\right)\right]}{\partial\tilde{T}}\right\}.\label{heatphi}
\end{eqnarray}
Since all important thermodynamic quantities for the reaction dynamics can be calculated from the potential $\varphi[\tilde{T},\tilde{\mu};x]$, we will use it, instead of the entropy density $\sigma\left[\epsilon,n,x\right]$, hereafter. 

Before closing this section, we consider the equilibrium state of the slow reaction dynamics. 
Owing to the second law, candidates of the equilibrium state are given by the variational form: 
\begin{equation}
\displaystyle \xi_{\mathrm{E}\mathrm{Q}}\in\arg\max_{\xi}\left\{-\varphi\left[\tilde{T},\tilde{\mu};x_{0}+S\xi\right]-\tilde{\mu}_{m}O_{r}^{m}\xi^{r}\right\},\label{VPXi}
\end{equation}
and $x_{\mathrm{E}\mathrm{Q}}=x_{0}+S\xi_{\mathrm{E}\mathrm{Q}}$. 
However, differently from the case in the fast dynamics (see Eq.(\ref{VPEN})), the existence and uniqueness of the equilibrium state are not guaranteed in this case because of $S$ and $O$. 
In the following sections, we will analyze the equilibrium state from a geometric point of view. 

\section{III. A geometric representation of equilibrium states}

In this section, we consider a geometric interpretation of the variational form, Eq. (\ref{VPXi}).
As a result, we reveal the existence and uniqueness condition for the equilibrium state. 

The geometry we use here is Hessian geometry \cite{c6}, which is based on a pair of linearly dual spaces. These spaces are endowed with a second dual structure resulting from Legendre transformation with a given convex function. The two dualities yield a generalized orthogonality relation between affine subspaces in the two spaces. 
Also, the convex function induces the Bregman divergence, which works as an asymmetric distance on the dual spaces.

As we demonstrate, Hessian geometry quite naturally captures the duality between chemical densities and chemical potentials linked by the thermodynamic convex function, and disentangle the algebraic constraints imposed by the stoichiometry of CRNs. 

\subsection{A. Preparation for geometry}
\begin{figure}[h]
\begin{center}
\includegraphics[width=0.45\textwidth]{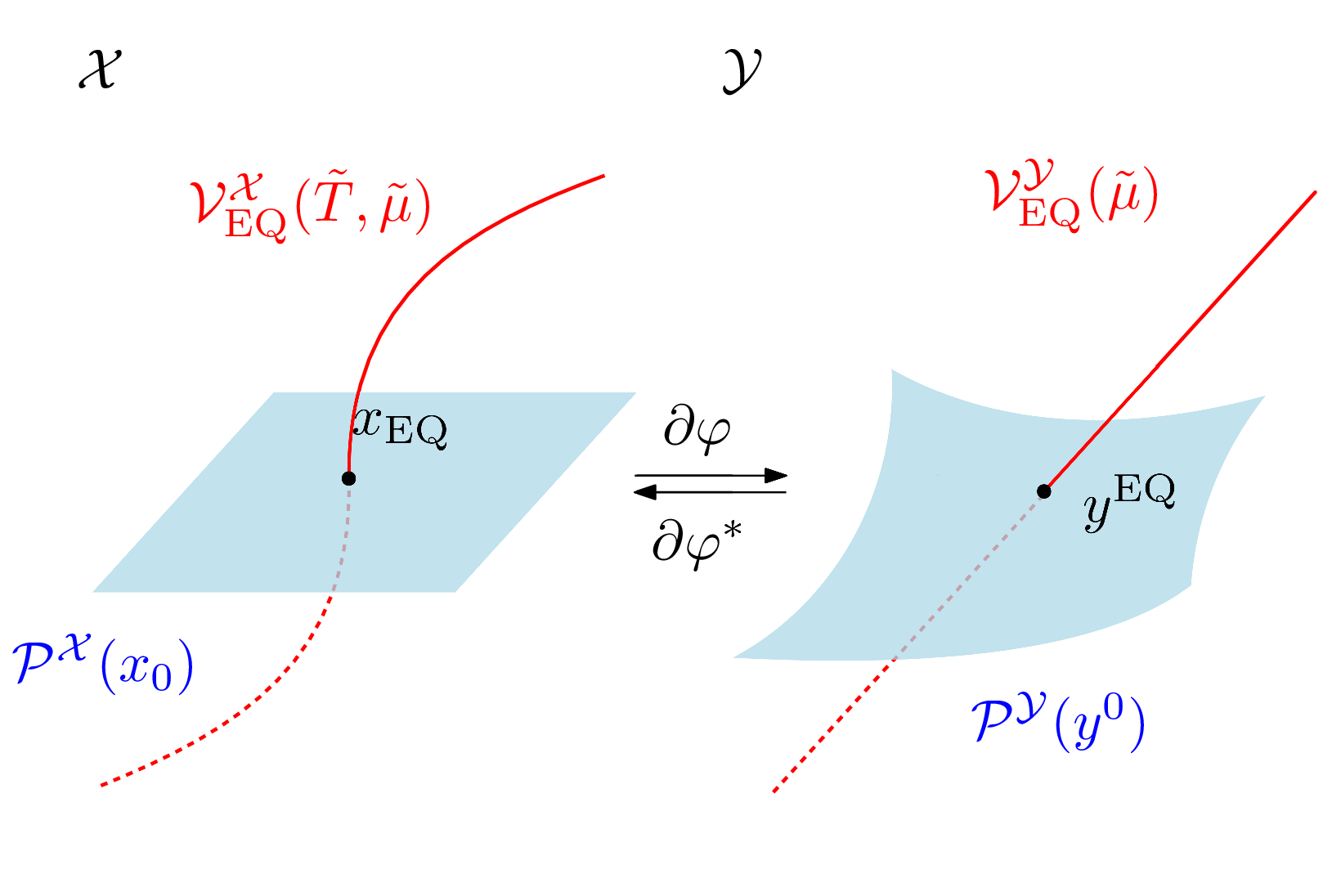}
\caption{The left and right spaces represent the density space $\mathcal{X}$ and the chemical potential space $\mathcal{Y}$, which are mapped each other by $\partial \varphi$ and $\partial \varphi^*$. The red manifold represents the equilibrium manifold, which is curved in $\mathcal{X}$ and is flat in $\mathcal{Y}$. By contrast, the blue manifold denotes the stoichiometric manifold, which is flat in $\mathcal{X}$ and is curved in $\mathcal{Y}$. The intersection between these two submanifolds gives the equilibrium state.
}
\label{fig2}
\end{center}
\end{figure}

We write $x\in \mathcal{X}=\mathbb{R}_{>0}^{\mathcal{N}_X}$ for the density space of the confined chemicals, where $\mathcal{N}_X$ is the number of species. 
Also, we define its dual space: $y\in \mathcal{Y}=\mathbb{R}^{\mathcal{N}_X}$, which is the corresponding chemical potential space. 
Consider a map from $\mathcal{X}$ to $\mathcal{Y}$ by using the convex function $\varphi\left(x\right):=\varphi[\tilde{T},\tilde{\mu};x]$ as 
\begin{equation}
\partial\varphi: x\in \mathcal{X}\mapsto\partial\varphi\left(x\right)=\left\{\partial_{i}\varphi\right\}=\left\{\frac{\partial\varphi}{\partial x^{i}}\right\}\in \mathcal{Y},\label{mapXtoY}
\end{equation}
where, to focus on $x$, we omit the arguments $\tilde{T}$ and $\tilde{\mu}$ in $\varphi$, and the convexity of $\varphi\left(x\right)$ is guaranteed by the definitions of thermodynamic potentials, Eqs. (\ref{TPm}), (\ref{TPf}) and (\ref{TPg}). 
In physical interpretation, the map $\partial\varphi$ gives the value of the chemical potential of a state $x$. 
Since we have assumed strict concavity for $\sigma$, which implies strict convexity of $\varphi\left(x\right)$, the map $\partial\varphi$ is injective. 
Furthermore, in the ordinary setting of chemical reaction systems, the range of $\partial\varphi$ is $\mathbb{R}^{\mathcal{N}_X}$; thus $\partial\varphi$ is bijective (see also \cite{n2}). 
To construct the inverse map of $\partial\varphi$, we define the strictly convex function $\varphi^{*}\left(y\right)$ on the dual space $\mathcal{Y}$ by the Legendre transformation: 
\begin{equation}
\displaystyle \varphi^{*}\left(y\right):=\max_{x}\left\{y_{i}x^{i}-\varphi\left(x\right)\right\}.\label{LTphi}
\end{equation}
Employing $\varphi^{*}\left(y\right)$, we can represent the inverse map as 
\begin{equation}
\partial\varphi^{*}: y\in \mathcal{Y}\mapsto\partial\varphi^{*}\left(y\right)=\left\{\partial^{i}\varphi^{*}\right\}=\left\{\frac{\partial\varphi^{*}}{\partial y_{i}}\right\}\in \mathcal{X}.\label{mapYtoX}
\end{equation}
The diagrammatic summary of these spaces and maps is shown in FIG. \ref{fig2}. 

With the above setup, we analyze the equilibrium state given by Eq. (\ref{VPXi}). 
The critical equation of the variational form, Eq. (\ref{VPXi}), is represented as 
\begin{equation}
A_{r}\left(\tilde{T},\tilde{\mu};x_{0}+S\xi\right):=-\partial_{i}\varphi\left(x_{0}+S\xi\right)S_{r}^{i}-\tilde{\mu}_{m}O_{r}^{m}=0,\label{CEQ}
\end{equation}
where we define the affinity $A(\tilde{T},\tilde{\mu};x)$ \cite{n5}. This measures how far a state $x$ is from the equilibrium state \cite{06,07}. 
The solutions of Eq. (\ref{CEQ}) with respect to $\xi$ give candidates of the equilibrium extent of reaction, $\xi_{\mathrm{E}\mathrm{Q}}$. 

Since it is difficult to directly analyze Eq. (\ref{CEQ}), we introduce its geometric representation. 
Define the following two submanifolds (subsets) in the density space $\mathcal{X}$ (see FIG. \ref{fig2}). 
One is the equilibrium manifold: 
\begin{equation}
\mathcal{V}_{\mathrm{E}\mathrm{Q}}^{\mathcal{X}}\left(\tilde{T},\tilde{\mu}\right):=\left\{x|A\left(\tilde{T},\tilde{\mu};x\right)=0\right\},
\end{equation}
which represents a set of candidates of the equilibrium state. 
The other is the stoichiometric manifold: 
\begin{equation}
\mathcal{P}^{\mathcal{X}}\left(x_{0}\right):=\left\{x|x\in x_{0}+{\rm Im}\left[S\right]\right\},\label{SmaniinX}
\end{equation}
which describes an affine subspace in $\mathcal{X}$ and expresses the domain in which the system can evolve by the reaction dynamics with an initial condition $x_{0}$ \cite{m1,m4,m5,m6}. 
The important points here are that the equilibrium manifold $\mathcal{V}_{\mathrm{E}\mathrm{Q}}^{\mathcal{X}}(\tilde{T},\tilde{\mu})$ is determined by the reservoir condition $(\tilde{T},\tilde{\mu})$, whereas the stoichiometric manifold $\mathcal{P}^{\mathcal{X}}\left(x_{0}\right)$ is given by an  initial condition $x_{0}$. 

By using these two submanifolds, we can identify candidates of the equilibrium state $x_{\mathrm{E}\mathrm{Q}}=x_{0}+S\xi_{\mathrm{E}\mathrm{Q}}$ with the intersection between them (see FIG. \ref{fig2}): 
\begin{equation}
x_{\mathrm{E}\mathrm{Q}}\left(\tilde{T},\tilde{\mu};x_{0}\right)\in \mathcal{V}_{\mathrm{E}\mathrm{Q}}^{\mathcal{X}}\left(\tilde{T},\tilde{\mu}\right)\cap \mathcal{P}^{\mathcal{X}}\left(x_{0}\right).\label{intXsp}
\end{equation}
If the intersection consists of precisely one point, the equilibrium state is uniquely determined by the variational form, Eq. (\ref{VPXi}), under a given initial condition $x_{0}$ \cite{n4}. 
By contrast, if the intersection is empty, the equilibrium state does not exist. 
In the next subsection, we will derive the existence condition for the equilibrium state and prove its uniqueness. 

\subsection{B. Existence and uniqueness condition for the equilibrium state}
Let us begin with the derivation of the existence condition, which is composed of two steps: (1) finding the condition for $\mathcal{V}_{\mathrm{E}\mathrm{Q}}^{\mathcal{X}}(\tilde{T},\tilde{\mu})\neq\emptyset$ and (2) proving $\mathcal{V}_{\mathrm{E}\mathrm{Q}}^{\mathcal{X}}(\tilde{T},\tilde{\mu})\neq\emptyset \Rightarrow \mathcal{V}_{\mathrm{E}\mathrm{Q}}^{\mathcal{X}}(\tilde{T},\tilde{\mu})\cap \mathcal{P}^{\mathcal{X}}(x_{0})\neq\emptyset$. 
To obtain the condition for $\mathcal{V}_{\mathrm{E}\mathrm{Q}}^{\mathcal{X}}(\tilde{T},\tilde{\mu})\neq\emptyset$, we introduce the equilibrium manifold in the chemical potential space $\mathcal{Y}$ by using the map $\partial\varphi$: 
\begin{equation}
\mathcal{V}_{\mathrm{E}\mathrm{Q}}^{\mathcal{Y}}\left(\tilde{\mu}\right):=\partial\varphi\left(\mathcal{V}_{\mathrm{E}\mathrm{Q}}^{\mathcal{X}}\right)=\left\{y|y_{i}S_{r}^{i}+\tilde{\mu}_{m}O_{r}^{m}=0\right\},
\end{equation}
which defines an affine subspace in $\mathcal{Y}$ (see FIG. \ref{fig2}) \cite{na2}. 
For $\mathcal{V}_{\mathrm{E}\mathrm{Q}}^{\mathcal{Y}}(\tilde{\mu})\neq\emptyset$, the simultaneous equations $y_{i}S_{r}^{i}+\tilde{\mu}_{m}O_{r}^{m}=0$ must be consistent.
The consistency condition is known as $O^{T}\tilde{\bm{\mu}}\in{\rm Im}[S^{T}]$ in linear algebra, where $\left(\cdot\right)^{T}$ denotes the transpose operation and $\tilde{\bm{\mu}}:=\left(\tilde{\mu}_{1},\tilde{\mu}_{2},...\right)^{T}$. 
If we use complete basis of ${\rm Ker}\left[S\right]$: $\left(\bm{V}_{1},\bm{V}_{2},...\right)=:\left\{V_{c}^{r}\right\}$ (i.e. $SV=0$), the consistency condition can be rewritten as, for all $c$,  
\begin{equation}
\tilde{\mu}_{m}O_{r}^{m}V_{c}^{r}=0,\label{esposi}
\end{equation}
where we employ ${\rm Im}[S^{T}]\perp{\rm Ker}\left[S\right]$.
The basis vectors $\{\bm{V}_{c}\}$ are called reaction cycles \cite{03,06,07,08,09} and the condition, Eq. (\ref{esposi}), says that all affinities along reaction cycles vanish, that is $A_{r}V_{c}^{r}=0$ for all $c$. 
In other words, $\tilde{\mu}_mO^m_rV^r_c$ represent chemical gradients in the reservoir \cite{06,07}; Eq. (\ref{esposi}) argues that the system does not feel external chemical gradients, and therefore the existence of the equilibrium state is expected. 
If and only if the condition, Eq. (\ref{esposi}), is satisfied, we obtain $\mathcal{V}_{\mathrm{E}\mathrm{Q}}^{\mathcal{X}}(\tilde{T},\tilde{\mu})\neq\emptyset$ because the inverse map $\partial\varphi^{*}$ exists and $\mathcal{V}_{\mathrm{E}\mathrm{Q}}^{\mathcal{X}}(\tilde{T},\tilde{\mu})=\partial\varphi^{*}(\mathcal{V}_{\mathrm{E}\mathrm{Q}}^{\mathcal{Y}})$. 

We proceed to the second step: $\mathcal{V}_{\mathrm{E}\mathrm{Q}}^{\mathcal{X}}(\tilde{T},\tilde{\mu})\neq\emptyset \Rightarrow \mathcal{V}_{\mathrm{E}\mathrm{Q}}^{\mathcal{X}}(\tilde{T},\tilde{\mu})\cap \mathcal{P}^{\mathcal{X}}\left(x_{0}\right)\neq\emptyset$. 
If the consistency condition, Eq. (\ref{esposi}), holds, the simultaneous equations $y_{i}S_{r}^{i}+\tilde{\mu}_{m}O_{r}^{m}=0$ have solutions. 
Denoting a particular solution by $y^{P}=\left\{y_{i}^{P}\right\}$, we get $\tilde{\mu}_{m}O_{r}^{m}=-y_{i}^{P}S_{r}^{i}$. 
The substitution of it into Eq. (\ref{VPXi}) leads to 
\begin{equation}
\displaystyle \xi_{\mathrm{E}\mathrm{Q}}\in\arg\max_{\xi}\left\{y_{i}^{P}S_{r}^{i}\xi^{r}-\varphi\left[\tilde{T},\tilde{\mu};x_{0}+S\xi\right]\right\}.
\end{equation}
By using the argument change $ x=x_{0}+S\xi$, we obtain
\begin{equation}
x_{\mathrm{E}\mathrm{Q}}\displaystyle \in\arg\max_{x\in \mathcal{P}^{\mathcal{X}}\left(x_{0}\right)}\left\{y_{i}^{P}x^{i}-\varphi\left[\tilde{T},\tilde{\mu};x\right]\right\}.\label{XEQVP}
\end{equation}
Here, we note that the function maximized in Eq. (\ref{XEQVP}) is bounded above on the density space $\mathcal{X}$ \cite{n6}, which means that the function is also bounded above on the affine subspace $\mathcal{P}^{\mathcal{X}}(x_{0})$.  
Thus, the equilibrium state $x_{\mathrm{E}\mathrm{Q}}$ must exist, that is, $\mathcal{V}_{\mathrm{E}\mathrm{Q}}^{\mathcal{X}}(\tilde{T},\tilde{\mu})\cap \mathcal{P}^{\mathcal{X}}\left(x_{0}\right)\neq\emptyset$. 

Combining the above two steps, we obtain the following theorem: 
\begin{thm}
Equilibrium states exist, if and only if the consistency condition Eq. (\ref{esposi}) is satisfied. 
In that case, the intersection between the equilibrium and stoichiometric manifolds (Eq. (\ref{intXsp}) or Eq. (\ref{VPXi})) is not empty.
\end{thm}
An analogous theorem was originally stated by Wegscheider \cite{c4,m3} and has been recently reported in Refs. \cite{06,07}, under the ideal gas assumption and mass action kinetics. 
Therefore, our theorem is a generalization of their statement because we use neither ideal gas nor mass action kinetics assumptions. 
Also, if the stoichiometric matrices $S$ and $O$ satisfy $OV=0$ (i.e., ${\rm Im}[O^{T}]\subset{\rm Im}[S^{T}]$), the system is a so-called unconditionally equilibrium system \cite{06,07}. 
This means that, for any choice of reservoir condition $(\tilde{T},\tilde{\mu})$, the system must converge to an equilibrium state. 

Next, we show uniqueness of the equilibrium state under a given initial condition $x_{0}$. 
Since the function maximized in Eq. (\ref{XEQVP}) is strictly concave and also bounded above on the affine subspace $\mathcal{P}^{\mathcal{X}}\left(x_{0}\right)$, the point $\ x_{\mathrm{E}\mathrm{Q}}$ is uniquely determined by Eq. (\ref{XEQVP}). 
Hence, we obtain the following theorem: 
\begin{thm}
If the consistency condition, Eq. (\ref{esposi}), is satisfied, under a given initial condition $x_{0}$, the system converges to the equilibrium state that is uniquely determined by Eq. (\ref{XEQVP}), 
that is, the intersection, Eq. (\ref{intXsp}), consists of precisely one point. 
\end{thm}
This theorem is a generalization of the Horn-Jackson theory for detailed-balanced CRNs \cite{m1,05,06,07}, which was more recently rephrased as Birch's theorem in the language of algebraic geometry \cite{m4,m5,m6}. 
As will be shown in Sec. V, if we assume ideal gas conditions, this theorem reduces to Birch's theorem. 

In the derivation of the theorems, one may be concerned with the arbitrariness in choosing a particular solution. However, even if we choose another particular solution, the derived equilibrium state $x_{\mathrm{E}\mathrm{Q}}$ is unchanged because a particular solution is just a reference point for $\mathcal{V}^{\mathcal{Y}}_{\mathrm{EQ}}(\tilde{\mu})$. 
That is, the choice of a particular solution amounts to fixing a ``gauge" in the theory. 

Finally, we comment on the equilibrium state in the chemical potential space $\mathcal{Y}$. 
On the one hand, by denoting a particular solution of the simultaneous equations $y_{i}S_{r}^{i}+\tilde{\mu}_{m}O_{r}^{m}=0$ by $y^{P}$, the general solution can be represented as
\begin{equation}
    y_{i}=y_{i}^{P}+\eta_{l}U_{i}^{l},
\end{equation}
where $\eta=\left\{\eta_{l}\right\}$ represents coordinates on ${\rm Ker}[S^{T}]$; Here, $U$ is a basis matrix: $\left\{U_{i}^{l}\right\}:=\left(\bm{U}^{1},\bm{U}^{2},...\right)^{T}$ whose rows $\bm{U}^{l}$ form a basis of ${\rm Ker}\left[S^{T}\right]$ (i.e. $US=0$).
Thus, we get a parameter representation of the equilibrium manifold $\mathcal{V}_{\mathrm{E}\mathrm{Q}}^{\mathcal{Y}}(\tilde{\mu})$ as 
\begin{equation}
\mathcal{V}_{\mathrm{E}\mathrm{Q}}^{\mathcal{Y}}\left(\tilde{\mu}\right)=\left\{y|y_{i}=y_{i}^{P}+\eta_{l}U_{i}^{l}, \eta_{l}\in \mathbb{R}\right\}.\label{VYU}
\end{equation}
On the other hand, by using $\partial\varphi$, we can map the stoichiometric manifold $\mathcal{P}^{\mathcal{X}}\left(x_{0}\right)$ into $\mathcal{Y}$: 
\begin{equation}
\mathcal{P}^{\mathcal{Y}}\left(y^{0}\right):=\partial\varphi\left(\mathcal{P}^{\mathcal{X}}\left(x_{0}\right)\right),
\end{equation}
where $y^{0}=\left\{y_{i}^{0}\right\}=\partial\varphi\left(x_{0}\right)$ represent the chemical potential at the initial state $x_{0}$ \cite{na2}. 
Here, we note that this manifold no longer describes an affine subspace in $\mathcal{Y}$, but a curved one in general (see FIG. \ref{fig2}). 
By employing the above two submanifolds, $\mathcal{V}_{\mathrm{E}\mathrm{Q}}^{\mathcal{Y}}(\tilde{\mu})$ and $\mathcal{P}^{\mathcal{Y}}\left(y^{0}\right)$, the equilibrium state can be characterized in $\mathcal{Y}$ as 
\begin{equation}
y^{\mathrm{E}\mathrm{Q}}=\left\{y_{i}^{\mathrm{E}\mathrm{Q}}\right\}=\partial\varphi\left(x_{\mathrm{E}\mathrm{Q}}\right)\in \mathcal{V}_{\mathrm{E}\mathrm{Q}}^{\mathcal{Y}}\left(\tilde{\mu}\right)\cap \mathcal{P}^{\mathcal{Y}}\left(y^{0}\right).
\end{equation}

\section{IV. The second law as minimization of divergence}
If the consistency condition, Eq. (\ref{esposi}), is satisfied, the time evolution of the total entropy, Eq. (\ref{enttotXi}), on the stoichiometric manifold $\mathcal{P}^{\mathcal{X}}\left(x_{0}\right)$ can be written as follows. 
By using $\tilde{\mu}_{m}O_{r}^{m}=-y_{i}^{P}S_{r}^{i}$, we get 
\begin{eqnarray}
\nonumber \displaystyle \Sigma^{\mathrm{t}\mathrm{o}\mathrm{t}}\left(x\left(t\right)\right)&=&\frac{\Omega}{\tilde{T}}\left\{y_{i}^{P}\left(x^{i}\left(t\right)-x^i_0\right)-\varphi\left[\tilde{T},\tilde{\mu};x\left(t\right)\right]\right\}\\&&+\mathrm{const}.,\label{prepent}
\end{eqnarray}
where $y^{P}\in \mathcal{V}_{\mathrm{E}\mathrm{Q}}^{\mathcal{Y}}(\tilde{\mu})$ and $x\left(t\right)\in \mathcal{P}^{\mathcal{X}}\left(x_{0}\right)$. 
We also note that the form of Eq. (\ref{prepent}) does not depend on the choice of a particular solution $y^{P}$ (see details in \cite{na}).
In this section, we give a geometric representation of Eq. (\ref{prepent}) through the Bregman divergence \cite{c6,g1,g2}. 
Moreover, we reformulate the second law from the viewpoint of Hessian geometry. As a result, we obtain four distinct characterizations of the equilibrium state; one of them is equivalent to Eq. (\ref{XEQVP}). 

\subsection{A. Entropy production during a relaxation to the equilibrium state}
\begin{figure}[h]
\begin{center}
\includegraphics[width=0.5\textwidth]{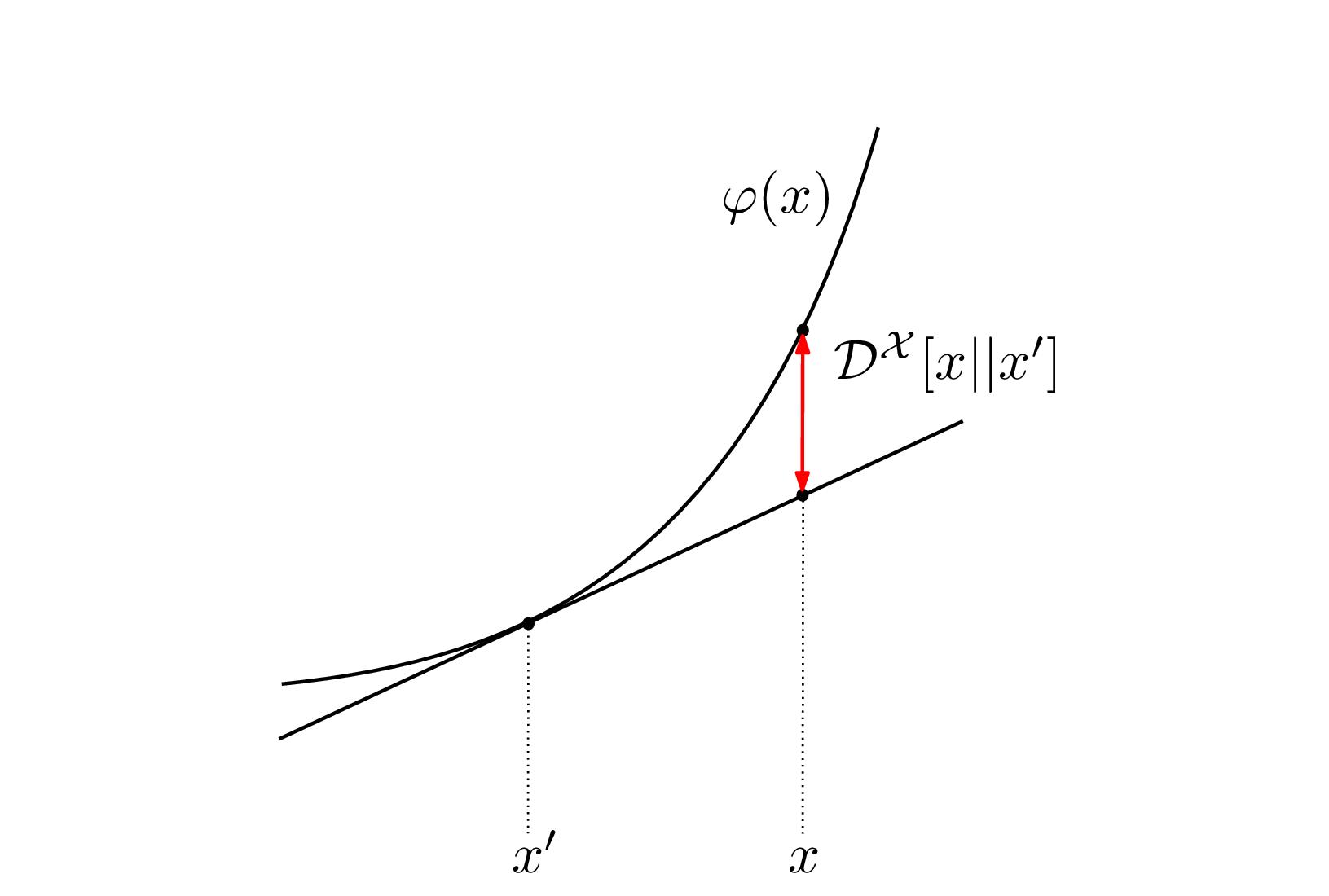}
\caption{The schematic definition of the Bregman divergence $\mathcal{D}^{\mathcal{X}}\left[x||x^{\prime}\right]$. The curve represents the convex function $\varphi(x)$. The line denotes the hyperplane tangent to $\varphi(\cdot)$ at the point $x'$. The divergence is given by the deviation between them at the point $x$, which is shown in red. }
\label{fig3}
\end{center}
\end{figure}

The Bregman divergence on $\mathcal{X}$ is defined by
\begin{equation}
\mathcal{D}^{\mathcal{X}}\left[x||x^{\prime}\right]:=\left\{\varphi\left(x\right)-\varphi\left(x^{\prime}\right)\right\}-\partial_{i}\varphi\left(x^{\prime}\right)\left\{x^{i}-\left(x^{\prime}\right)^{i}\right\}.\label{BDX}
\end{equation}
It measures the deviation at a point $x$ between the convex function $\varphi\left(x\right)$ and the hyperplane tangent to it at a point $x^{\prime}$ (see FIG. \ref{fig3}).
This divergence has the following property: $\mathcal{D}^{\mathcal{X}}\left[x||x^{\prime}\right]\geq 0$, and equality holds if and only if $x=x^{\prime}$, i.e., it acts as an asymmetric distance from $x^{\prime}$ to $x$. 
By employing the divergence, we can calculate the production (increment) of the total entropy, Eq. (\ref{prepent}), during a time interval $\left[t^{\prime},t\right]$ as 
\begin{eqnarray}
\nonumber&&\Sigma^{\mathrm{t}\mathrm{o}\mathrm{t}}\left(x\left(t\right)\right)-\Sigma^{\mathrm{t}\mathrm{o}\mathrm{t}}\left(x\left(t^{\prime}\right)\right)\\
&&=-\displaystyle \frac{\Omega}{\tilde{T}}\left\{\mathcal{D}^{\mathcal{X}}\left[x\left(t\right)||x_{P}\right]-\mathcal{D}^{\mathcal{X}}\left[x\left(t^{\prime}\right)||x_{P}\right]\right\},\label{III1}
\end{eqnarray}
where $x_{P}:=\partial\varphi^{*}\left(y^{P}\right)\in \mathcal{V}_{\mathrm{E}\mathrm{Q}}^{\mathcal{X}}(\tilde{T},\tilde{\mu})$. 
Here, we also used the fact that both $x(t)$ and $x(t')$ are on the stoichiometric manifold $\mathcal{P}^{\mathcal{X}}(x_0)$ to cancel out the term depending on the initial condition $x_0$ in Eq. (\ref{prepent}). 
Using this representation, we can evaluate the heat dissipation, Eq. (\ref{heatphi}), involving the divergence as 
\begin{eqnarray}
\nonumber&&\mathcal{Q}_{t^{\prime}\rightarrow t}=-\Omega\left\{\mathcal{D}^{\mathcal{X}}\left[x\left(t\right)||x_{P}\right]-\mathcal{D}^{\mathcal{X}}\left[x\left(t^{\prime}\right)||x_{P}\right]\right\}\\
&&+\Omega\tilde{T}\left\{\frac{\partial\varphi\left[\tilde{T},\tilde{\mu};x\left(t\right)\right]}{\partial\tilde{T}}-\frac{\partial\varphi\left[\tilde{T},\tilde{\mu};x\left(t^{\prime}\right)\right]}{\partial\tilde{T}}\right\}.
\end{eqnarray}
For a relaxation to the equilibrium state, the production of the total entropy, Eq. (\ref{III1}), is computed as 
\begin{eqnarray}
\nonumber&&\Sigma^{\mathrm{t}\mathrm{o}\mathrm{t}}\left(x_{\mathrm{E}\mathrm{Q}}\right)-\Sigma^{\mathrm{t}\mathrm{o}\mathrm{t}}\left(x_{0}\right)\\
\displaystyle \nonumber&&=-\frac{\Omega}{\tilde{T}}\left\{\mathcal{D}^{\mathcal{X}}\left[x_{\mathrm{E}\mathrm{Q}}||x_{P}\right]-\mathcal{D}^{\mathcal{X}}\left[x_{0}||x_{P}\right]\right\}=\frac{\Omega}{\tilde{T}}\mathcal{D}^{\mathcal{X}}\left[x_{0}||x_{\mathrm{E}\mathrm{Q}}\right],\\\label{III3}
\end{eqnarray}
where we choose the equilibrium state $x_{\mathrm{E}\mathrm{Q}}$ as a particular state $x_{P}\in \mathcal{V}_{\mathrm{E}\mathrm{Q}}^{\mathcal{X}}(\tilde{T},\tilde{\mu})$ in the second equality. 
Thus, the heat dissipation during the relaxation can be represented as 
\begin{eqnarray}
\nonumber&&\mathcal{Q}_{0\rightarrow \mathrm{E}\mathrm{Q}}=\Omega \mathcal{D}^{\mathcal{X}}\left[x_{0}||x_{\mathrm{E}\mathrm{Q}}\right]\\
&&+\Omega\tilde{T}\left\{\frac{\partial\varphi\left[\tilde{T},\tilde{\mu};x_{\mathrm{E}\mathrm{Q}}\right]}{\partial\tilde{T}}-\frac{\partial\varphi\left[\tilde{T},\tilde{\mu};x_{0}\right]}{\partial\tilde{T}}\right\}.\label{III5}
\end{eqnarray}

The above result can be summarized as follows: 
\begin{thm}
If a CRN relaxes to the equilibrium state (i.e., the consistency condition, Eq. (\ref{esposi}), is satisfied), then the total entropy production during a relaxation from an initial state $x_0$ to the corresponding equilibrium state $x_{\mathrm{EQ}}$ can be evaluated by the Bregman divergence given by Eq. (\ref{III3}). 
Furthermore, the heat dissipation during the relaxation is calculated by Eq. (\ref{III5}). 
\end{thm}
This theorem represents a generalization of the result by Rao and Esposito \cite{07}, which was also reported in the context of mass action systems in Refs. \cite{a3,a4,m1}.   
As shown in Sec. V, if we assume ideal gas conditions, the Bregman divergence reduces to the generalized Kullback-Leibler divergence, and our statement corresponds to their result. 

\subsection{B. Characterizations of the equilibrium state}
\begin{figure}[h]
\begin{center}
\includegraphics[width=0.5\textwidth]{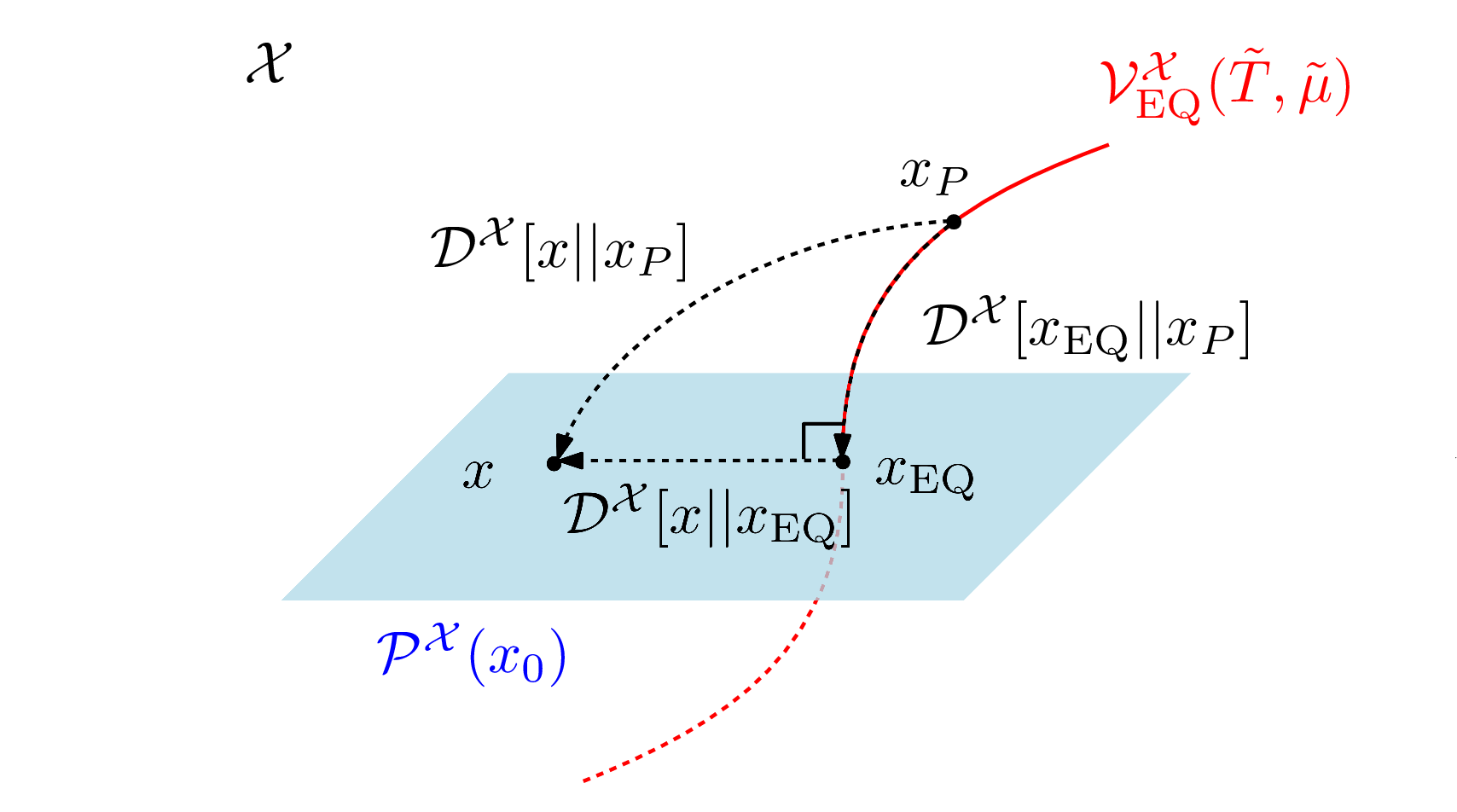}
\caption{The equilibrium and stoichiometric manifolds in $\mathcal{X}$ orthogonally intersect at the equilibrium state $x_{\mathrm{EQ}}$ \cite{n7}. 
The Pythagorean theorem states that the divergence of hypotenuse, $\mathcal{D}^{\mathcal{X}}\left[x||x_P \right]$, is equal to the sum of the ones on the other two sides, $\mathcal{D}^{\mathcal{X}}\left[x||x_{\mathrm{E}\mathrm{Q}} \right]$ and $\mathcal{D}^{\mathcal{X}}\left[x_{\mathrm{E}\mathrm{Q}}||x_P \right]$.  }
\label{fig4}
\end{center}
\end{figure}

Next, we characterize the equilibrium state by four distinct variational forms based on the divergence. 
For any three points, $x,\ x^{\prime}$ and $x^{\prime\prime}$ in $\mathcal{X}$, the following equality holds: 
\begin{eqnarray}
\nonumber&&\mathcal{D}^{\mathcal{X}}\left[x||x^{\prime}\right]+\mathcal{D}^{\mathcal{X}}\left[x^{\prime}||x^{\prime\prime}\right]\\
&&=\mathcal{D}^{\mathcal{X}}\left[x||x^{\prime\prime}\right]+\left\{x^{i}-\left(x^{\prime}\right)^{i}\right\}\left\{\partial_{i}\varphi\left(x^{\prime}\right)-\partial_{i}\varphi\left(x^{\prime\prime}\right)\right\}.\label{III2}
\end{eqnarray}
Since we have assumed that the consistency condition, Eq. (\ref{esposi}), holds, the equilibrium manifold is not empty, $\mathcal{V}_{\mathrm{E}\mathrm{Q}}^{\mathcal{X}}(\tilde{T},\tilde{\mu})\neq\emptyset$, and the unique equilibrium state $x_{\mathrm{E}\mathrm{Q}}\in \mathcal{V}_{\mathrm{E}\mathrm{Q}}^{\mathcal{X}}(\tilde{T},\tilde{\mu})\cap \mathcal{P}^{\mathcal{X}}(x_{0})$ exists. 
If we choose $x\in \mathcal{P}^{\mathcal{X}}\left(x_{0}\right),\ x^{\prime}=x_{\mathrm{E}\mathrm{Q}}$ and $x^{\prime\prime}=x_{P}\in \mathcal{V}_{\mathrm{E}\mathrm{Q}}^{\mathcal{X}}(\tilde{T},\tilde{\mu})$, the last term in the right hand side of Eq. (\ref{III2}) vanishes, because
\begin{eqnarray}
\nonumber&&\left\{x^{i}-x_{\mathrm{E}\mathrm{Q}}^{i}\right\}\left\{\partial_{i}\varphi\left(x_{\mathrm{E}\mathrm{Q}}\right)-\partial_{i}\varphi\left(x_{P}\right)\right\}\\
&&=\left\{x^{i}-x_{\mathrm{E}\mathrm{Q}}^{i}\right\}\left\{y_{i}^{\mathrm{E}\mathrm{Q}}-y_{i}^{P}\right\}=0,
\end{eqnarray}
where we use the facts that $x-x_{\mathrm{E}\mathrm{Q}}\in{\rm Im}\left[S\right],\ y^{\mathrm{E}\mathrm{Q}}-y^{P}\in{\rm Ker}\left[S^{T}\right]$ and ${\rm Im}\left[S\right]\perp{\rm Ker}\left[S^{T}\right]$. 
This represents the orthogonality between $\mathcal{P}^{\mathcal{X}}(x_{0})$ and $\mathcal{V}_{\mathrm{E}\mathrm{Q}}^{\mathcal{X}}(\tilde{T},\tilde{\mu})$ at $x_{\mathrm{E}\mathrm{Q}}$ \cite{c6,g1}. 
Thus, we get the generalized Pythagorean theorem (see FIG. \ref{fig4}): 
\begin{equation}
\mathcal{D}^{\mathcal{X}}\left[x||x_{\mathrm{E}\mathrm{Q}}\right]+\mathcal{D}^{\mathcal{X}}\left[x_{\mathrm{E}\mathrm{Q}}||x_{P}\right]=\mathcal{D}^{\mathcal{X}}\left[x||x_{P}\right].\label{PYT}
\end{equation}
From this equality, we can derive two distinct variational forms to characterize the equilibrium state $x_{\mathrm{E}\mathrm{Q}}$. 

First, we minimize Eq. (\ref{PYT}) with respect to $x$ in the stoichiometric manifold $\mathcal{P}^{\mathcal{X}}\left(x_{0}\right)$. Then, we obtain 
\begin{equation}
x_{\mathrm{E}\mathrm{Q}}=\displaystyle \arg\min_{x\in \mathcal{P}^{\mathcal{X}}\left(x_{0}\right)}\mathcal{D}^{\mathcal{X}}\left[x||x_{P}\right],\label{DVP1}
\end{equation}
where we use $x_{\mathrm{E}\mathrm{Q}}=\arg \mathrm{min}_{x\in \mathcal{P}^{\mathcal{X}}\left(x_{0}\right)}\mathcal{D}^{\mathcal{X}}\left[x||x_{\mathrm{E}\mathrm{Q}}\right]$. 
Taking Eqs. (\ref{prepent}) and (\ref{III1}) into account, we find that this variational form coincides with Eq. (\ref{XEQVP}); 
that is, Eq. (\ref{DVP1}) implies the conventional characterization of the equilibrium state by the second law. 

Second, if we minimize Eq. (\ref{PYT}) with respect to $x_{P}$ in the equilibrium manifold $\mathcal{V}_{\mathrm{E}\mathrm{Q}}^{\mathcal{X}}(\tilde{T},\tilde{\mu})$ and set $x=x_{0}$, we get another non-trivial variational form:
\begin{equation}
x_{\mathrm{E}\mathrm{Q}}=\displaystyle \arg\min_{x_{P}\in \mathcal{V}_{\mathrm{E}\mathrm{Q}}^{\mathcal{X}}\left(\tilde{T},\tilde{\mu}\right)}\mathcal{D}^{\mathcal{X}}\left[x_{0}||x_{P}\right],\label{DVP2a}
\end{equation}
where we use $x_{\mathrm{E}\mathrm{Q}}=\arg \mathrm{min}_{x_{P}\in \mathcal{V}_{\mathrm{E}\mathrm{Q}}^{\mathcal{X}}(\tilde{T},\tilde{\mu})}\mathcal{D}^{\mathcal{X}}\left[x_{\mathrm{E}\mathrm{Q}}||x_{P}\right]$. 
In addition, from Eq. (\ref{III3}), the total entropy production for a relaxation can be evaluated as
\begin{equation}
\displaystyle \Sigma^{\mathrm{t}\mathrm{o}\mathrm{t}}\left(x_{\mathrm{E}\mathrm{Q}}\right)-\Sigma^{\mathrm{t}\mathrm{o}\mathrm{t}}\left(x_{0}\right)=\frac{\Omega}{\tilde{T}}\min_{x_{P}\in \mathcal{V}_{\mathrm{E}\mathrm{Q}}^{\mathcal{X}}\left(\tilde{T},\tilde{\mu}\right)}\mathcal{D}^{\mathcal{X}}\left[x_{0}||x_{P}\right].\label{III4}
\end{equation}
Owing to Eq. (\ref{III4}), we can evaluate the heat dissipation during the relaxation by using Eq. (\ref{III5}). 

The above framework constructed in the density space $\mathcal{X}$ can be mapped to the chemical potential space $\mathcal{Y}$. 
We define the Bregman divergence on $\mathcal{Y}$ by 
\begin{equation}
\mathcal{D}^{\mathcal{Y}}\left[y||y^{\prime}\right]:=\left\{\varphi^{*}\left(y\right)-\varphi^{*}\left(y^{\prime}\right)\right\}-\partial^{i}\varphi^{*}\left(y^{\prime}\right)\left\{y_{i}-y_{i}^{\prime}\right\}.\label{BDY}
\end{equation}
Since the equality, $\mathcal{D}^{\mathcal{X}}\left[x||x^{\prime}\right]=\mathcal{D}^{\mathcal{Y}}\left[\partial\varphi\left(x^{\prime}\right)||\partial\varphi\left(x\right)\right]$ holds, we get the Pythagorean theorem in $\mathcal{Y}$ as 
\begin{equation}
\mathcal{D}^{\mathcal{Y}}\left[y^{P}||y^{\mathrm{E}\mathrm{Q}}\right]+\mathcal{D}^{\mathcal{Y}}\left[y^{\mathrm{E}\mathrm{Q}}||y\right]=\mathcal{D}^{\mathcal{Y}}\left[y^{P}||y\right],\label{PYTY}
\end{equation}
where $y^{P}\in \mathcal{V}_{\mathrm{E}\mathrm{Q}}^{\mathcal{Y}}(\tilde{\mu}),\ y\in \mathcal{P}^{\mathcal{Y}}\left(y^{0}\right)$ and $y^{\mathrm{E}\mathrm{Q}}=\partial\varphi\left(x_{\mathrm{E}\mathrm{Q}}\right)$;  $y^{0}=\partial\varphi\left(x_{0}\right)$. 
By employing the same discussion as for the density space $\mathcal{X}$, the equality, Eq.  (\ref{PYTY}), yields the other two variational forms in $\mathcal{Y}$ to characterize the equilibrium state $y^{\mathrm{E}\mathrm{Q}}$. 
One is given by the minimization of Eq. (\ref{PYTY}) with respect to $y$ in $\mathcal{P}^{\mathcal{Y}}\left(y^{0}\right)$: 
\begin{equation}
y^{\mathrm{E}\mathrm{Q}}=\displaystyle \arg\min_{y\in \mathcal{P}^{\mathcal{Y}}\left(y^{0}\right)}\mathcal{D}^{\mathcal{Y}}\left[y^{P}||y\right],\label{DVP3}
\end{equation}
which corresponds to Eq. (\ref{DVP1}). 
The other is obtained by the minimization of Eq. (\ref{PYTY}) with respect to $y^{P}$ in $\mathcal{V}_{\mathrm{E}\mathrm{Q}}^{\mathcal{Y}}(\tilde{\mu})$:
\begin{eqnarray}
&&y^{\mathrm{E}\mathrm{Q}}=\displaystyle \arg\min_{y^{P}\in \mathcal{V}_{\mathrm{E}\mathrm{Q}}^{\mathcal{Y}}\left(\tilde{\mu}\right)}\mathcal{D}^{\mathcal{Y}}\left[y^{P}||y^{0}\right],\label{main1}\\
\displaystyle \nonumber&&\Sigma^{\mathrm{t}\mathrm{o}\mathrm{t}}\left(y^{\mathrm{E}\mathrm{Q}}\right)-\Sigma^{\mathrm{t}\mathrm{o}\mathrm{t}}\left(y^{0}\right)=\frac{\Omega}{\tilde{T}}\min_{y^{P}\in \mathcal{V}_{\mathrm{E}\mathrm{Q}}^{\mathcal{Y}}\left(\tilde{\mu}\right)}\mathcal{D}^{\mathcal{Y}}\left[y^{P}||y^{0}\right],\\\label{main2}
\end{eqnarray}
which correspond to Eqs. (\ref{DVP2a}) and (\ref{III4}). 
Of course, from these variational forms, Eqs. (\ref{main1}) and (\ref{main2}), we can evaluate the heat dissipation during the relaxation as
\begin{eqnarray}
\nonumber&&\mathcal{Q}_{0\rightarrow \mathrm{E}\mathrm{Q}}=\Omega \mathcal{D}^{\mathcal{Y}}\left[y^{\mathrm{E}\mathrm{Q}}||y^{0}\right]\\
&&+\Omega\tilde{T}\left\{\frac{\partial\varphi^{*}\left[\tilde{T},\tilde{\mu},y^{\mathrm{E}\mathrm{Q}}\right]}{\partial\tilde{T}}-\frac{\partial\varphi^{*}\left[\tilde{T},\tilde{\mu},y^{0}\right]}{\partial\tilde{T}}\right\},\label{heaty}
\end{eqnarray}
where we write all arguments of $\varphi^{*}\left(y\right)$ as $\varphi^{*}[\tilde{T},\tilde{\mu},y]$. 

\begin{figure}[h]
\begin{center}
\includegraphics[width=0.5\textwidth]{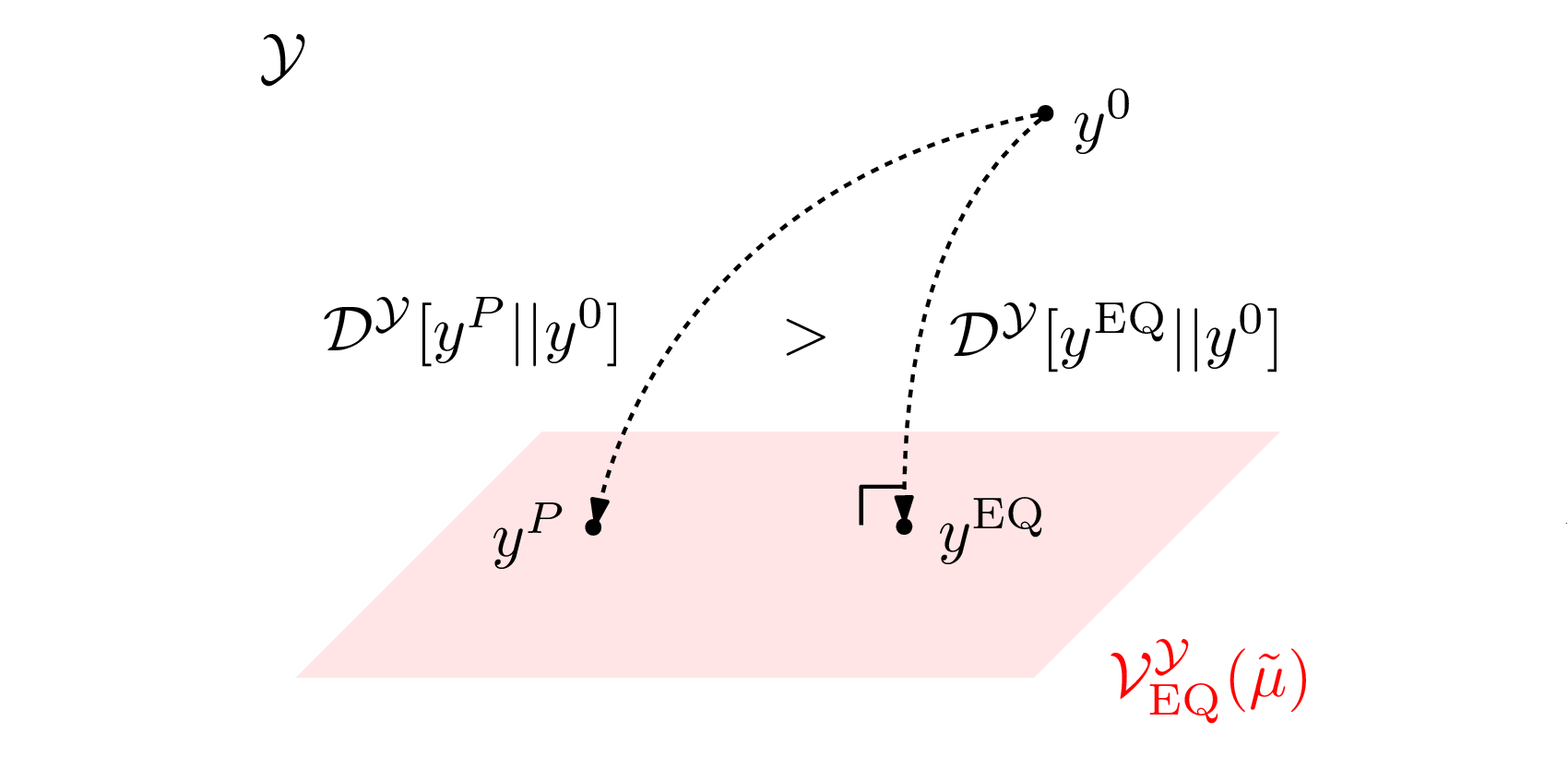}
\caption{The initial state in $\mathcal{Y}$ is denoted by $y^0$. The total entropy production is given by the minimized divergence from $y^0$ to the equilibrium manifold $\mathcal{V}_{\mathrm{E}\mathrm{Q}}^{\mathcal{Y}}(\tilde{\mu})$, which is an affine subspace in $\mathcal{Y}$. The orthogonal projection of $y^0$ to $\mathcal{V}_{\mathrm{E}\mathrm{Q}}^{\mathcal{Y}}(\tilde{\mu})$ represents the equilibrium state $y^{\mathrm{EQ}}$. }
\label{fig5}
\end{center}
\end{figure}

The above four characterizations of the equilibrium state are the main results of this work, which is summarized as follows: 
\begin{thm}
Consider a CRN such that the stoichiometric matrices $S$ and $O$ satisfy the consistency condition, Eq. (\ref{esposi}) (i.e. the CRN relaxes to the equilibrium state). 
Define the Bregman divergences in the density space $\mathcal{X}$ and the chemical potential spaces $\mathcal{Y}$ by Eqs. (\ref{BDX}) and (\ref{BDY}), respectively; 
the convex function $\varphi(x)$ represents the partial grand potential density given by Eq. (\ref{TPg}), and $\varphi^*(y)$ is its Legendre dual function as in Eq. (\ref{LTphi}). 
Then, in the density space $\mathcal{X}$, the equilibrium state $x_{\mathrm{EQ}}$ for a given initial state $x_0$ is characterized by the two distinct variational forms, Eqs. (\ref{DVP1}) and (\ref{DVP2a}).  
Also, the total entropy production during a relaxation to $x_{\mathrm{EQ}}$ is evaluated by Eq. (\ref{III4}).  
Furthermore, in the chemical potential space $\mathcal{Y}$, the equilibrium state $y^{\mathrm{EQ}}$ for a given initial state $y^0$ is determined by the other two distinct variational forms, Eqs. (\ref{DVP3}) and (\ref{main1});  
the total entropy production during a relaxation to $y^{\mathrm{EQ}}$ is computed by Eq. (\ref{main2}). 
\end{thm}
In particular, the variational forms, Eqs. (\ref{main1}) and (\ref{main2}), lead us to the following simple prescription to identify the equilibrium state: 
\begin{remark}
$\left(\bm{0}\right)$ Confirm the consistency condition, Eq. (\ref{esposi}), from given stoichiometric matrices, $S$ and $O$, and the reservoir condition $(\tilde{T},\tilde{\mu})$. 
If it does not hold, the equilibrium state does not exist, and the total entropy is diverging in the time evolution. 
$\left(\bm{1}\right)$ Calculate $\varphi\left(x\right)$ and $\varphi^{*}\left(y\right)$ by the Legendre transformation of a given thermodynamic potential.
$\left(\bm{2}\right)$ Obtain the equilibrium manifold $\mathcal{V}_{\mathrm{E}\mathrm{Q}}^{\mathcal{Y}}(\tilde{\mu})$ in chemical potential space $\mathcal{Y}$ as in Eq. (\ref{VYU}) by solving the simultaneous equations $y_{i}S_{r}^{i}+\tilde{\mu}_{m}O_{r}^{m}=0$. 
$\left(\bm{3}\right)$ Compute the corresponding chemical potential $y^{0}$ by applying the map $\partial\varphi$ to a given initial condition $x_{0}$. 
$\left(\bm{4}\right)$ Obtain the equilibrium state $y^{\mathrm{E}\mathrm{Q}}$ in the minimization problem of the divergence, i.e., the variational form, Eq. (\ref{main1}).
$\left(\bm{5}\right)$ If one wants to know the equilibrium density of confined chemicals, $x_{\mathrm{E}\mathrm{Q}}$, it is given by using the inverse map $\partial\varphi^{*}$. 
Also, the heat dissipation during the relaxation is computed by Eqs. (\ref{main2}) and (\ref{heaty}). 
\end{remark}
The schematic explanation of Prescription 1 is shown in FIG. \ref{fig5}. 

\section{V. Connection to previous work}
In the preceding sections, we have not imposed detailed functional forms on the thermodynamic potential or the flux density. We have only assumed for them that the potential is a convex (or a concave) function and the flux density satisfies the second law, 
which guarantee the increasing property of the total entropy function $\Sigma^{\mathrm{t}\mathrm{o}\mathrm{t}}\left(\xi\left(t\right)\right)$. 
In this section, we take the ideal gas potential and the mass action kinetics as specific forms of the thermodynamic potential $\varphi[\tilde{T},\tilde{\mu};x]$ and the reaction flux density $j\left(t\right)$.
As a result, a connection to previous work is clarified. 

Readers, who are familiar with kinetic modeling of CRNs, can refer to our accompanying paper \cite{KobaAccompaning}.
There, we derive the results of this section starting from the mass action kinetics and detailed balancing.

\subsection{A. Ideal gas}
In this subsection, under the ideal gas assumption, we demonstrate the geometric structure of thermodynamics constructed in the preceding sections. 
The form of the Helmholtz free-energy density for the ideal gas is known as 
\begin{eqnarray}
\displaystyle \nonumber f\left[\tilde{T};n,x\right]&=&n^{m}\displaystyle \mu_{m}^{o}\left(\tilde{T}\right)+R\tilde{T}\sum_{m}\left\{n^{m}\log n^{m}-n^{m}\right\}\\
&&+x^{i}\displaystyle \nu_{i}^{o}\left(\tilde{T}\right)+R\tilde{T}\sum_{i}\left\{x^{i}\log x^{i}-x^{i}\right\},\label{IDHF}
\end{eqnarray}
where $R$ represents the gas constant; $\mu^{o}(\tilde{T})=\{\mu_{m}^{o}(\tilde{T})\}$ and $\nu^{o}(\tilde{T})=\{\nu_{i}^{o}(\tilde{T})\}$ denote the standard chemical potentials of open and confined chemicals, which are functions only of the temperature $\tilde{T}$. 
Details of the definitions of $\mu^{o}(\tilde{T})$ and $\nu^{o}(\tilde{T})$ are shown in Appendix B. 
By using Eq. (\ref{TPg}), we can calculate the partial grand potential density as 
\begin{eqnarray}
\displaystyle \nonumber\varphi\left[\tilde{T},\tilde{\mu};x\right]=\varphi\left(x\right)&=&x^{i}\displaystyle \nu_{i}^{o}\left(\tilde{T}\right)+R\tilde{T}\sum_{i}\left\{x^{i}\log x^{i}-x^{i}\right\}\\
&&-R\displaystyle \tilde{T}\sum_{m}e^{\left\{\tilde{\mu}_{m}-\mu_{m}^{o}\left(\tilde{T}\right)\right\}/R\tilde{T}}.\label{IGTPg}
\end{eqnarray}

Since the differentiation of $\varphi[\tilde{T},\tilde{\mu};x]$ with respect to $\tilde{\mu}$ leads to the quasi-equilibrium density of the open chemicals $n_{\mathrm{Q}\mathrm{E}\mathrm{Q}}(\tilde{T},\tilde{\mu};x)$ (see Eq. (\ref{nQEQ}) in Appendix A), we get 
\begin{equation}
n_{\mathrm{Q}\mathrm{E}\mathrm{Q}}^{m}\displaystyle \left(\tilde{T},\tilde{\mu};x\right)=-\frac{\partial\varphi\left(\tilde{T},\tilde{\mu};x\right)}{\partial\tilde{\mu}_{m}}=e^{\left\{\tilde{\mu}_{m}-\mu_{m}^{o}\left(\tilde{T}\right)\right\}/R\tilde{T}}.\label{nqeqIG}
\end{equation}
This argues that, under a given constant $\tilde{\mu}$, the density of the open chemicals is kept to be constant in the reaction dynamics for the ideal gas cases. 
This is a natural consequence, because the ideal gas does not have any interactions among chemicals. 
Furthermore, if the reservoir also consists of ideal gas, its chemical potentials can be represented as 
\begin{equation}
\tilde{\mu}_{m}=\mu_{m}^{o}\left(\tilde{T}\right)+R\tilde{T}\log\tilde{n}^{m},
\end{equation}
where $\tilde{n}=\left\{\tilde{n}^{m}\right\}$ is the density of the open chemicals in the reservoir. 
Thus, the quasi-equilibrium density of the open chemicals in the system is equivalent to the one in the reservoir, $n_{\mathrm{Q}\mathrm{E}\mathrm{Q}}=\tilde{n}$. 
Also, the last term in Eq. (\ref{IGTPg}) becomes the total density of the open chemicals, $R\tilde{T}\Sigma_{m}\tilde{n}^{m}$. 

From Eq. (\ref{mapXtoY}), the map $\partial\varphi$ from $\mathcal{X}$ to $\mathcal{Y}$ is represented as
\begin{equation}
\partial_{i}\varphi\left(x\right)=\nu_{i}^{o}\left(\tilde{T}\right)+R\tilde{T}\log x^{i},\label{IGmapXtoY}
\end{equation}
which gives the chemical potential for the confined chemicals at a state $x$. 
The dual convex function $\varphi^{*}\left(y\right)$ on $\mathcal{Y}$ is calculated by the Legendre transformation, Eq. (\ref{LTphi}), as
\begin{eqnarray}
\displaystyle \nonumber\varphi^{*}\left(y\right)&=&R\displaystyle \tilde{T}\sum_{i}e^{\left\{y_{i}-\nu_{i}^{0}\left(\tilde{T}\right)\right\}/R\tilde{T}}\\
&&+R\displaystyle \tilde{T}\sum_{m}e^{\left\{\tilde{\mu}_{m}-\mu_{m}^{o}\left(\tilde{T}\right)\right\}/R\tilde{T}}.\label{IGDTPg}
\end{eqnarray}
Therefore, the map $\partial\varphi^{*}$ from $\mathcal{Y}$ to $\mathcal{X}$ is 
\begin{equation}
\partial^{i}\varphi^{*}\left(y\right)=e^{\left\{y_{i}-\nu_{i}^{0}\left(\tilde{T}\right)\right\}/R\tilde{T}},\label{IGmapYtoX}
\end{equation}
which is the inverse map of $\partial\varphi$. 

If the consistency condition, Eq. (\ref{esposi}), holds, by using the inverse map, Eq. (\ref{IGmapYtoX}), we get a parameter representation of the equilibrium manifold in $\mathcal{X}$ as 
\begin{equation}
\mathcal{V}_{\mathrm{E}\mathrm{Q}}^{\mathcal{X}}\left(\tilde{T},\tilde{\mu}\right)=\partial\varphi^{*}\left(\mathcal{V}_{\mathrm{E}\mathrm{Q}}^{\mathcal{Y}}\right)=\left\{x|x^{i}=\Gamma_{P}^{i}e^{\eta_{l}U_{i}^{l}}, \eta_{l}\in \mathbb{R}\right\},\label{toricTD}
\end{equation}
where we define $\Gamma_{P}^{i}:=e^{\left\{y_{i}^{P}-\nu_{i}^{0}\left(\tilde{T}\right)\right\}/R\tilde{T}}$. 
A submanifold expressed by the form of Eq. (\ref{toricTD}) is known as the exponential family or the toric model \cite{g1,g3}. 
In this algebro-geometric language, we can rephrase Theorem 2 in the ideal gas cases as 
\begin{thm} {\bf (}{\it {\bf Birch}}{\bf '}{\it {\bf s} {\bf Theorem}}{\bf )} 
The intersection between the toric model, Eq. (\ref{toricTD}), and the affine subspace (the stoichiometric manifold), Eq. (\ref{SmaniinX}), consists of precisely one point in the density space $\mathcal{X}$. 
\end{thm}
This theorem is known as Birch's theorem \cite{m4,m5,m6}, which is employed not only for chemical reaction systems but also for the maximum likelihood estimation in statistics \cite{g3}. 

Next, we calculate the Bregman divergences on $\mathcal{X}$ and $\mathcal{Y}$, Eqs. (\ref{BDX}) and (\ref{BDY}): 
\begin{eqnarray}
&&\displaystyle \mathcal{D}^{\mathcal{X}}\left[x||x^{\prime}\right]=R\tilde{T}\sum_{i}\left[x^{i}\log\frac{x^{i}}{\left(x^{\prime}\right)^{i}}-\left\{x^{i}-\left(x^{\prime}\right)^{i}\right\}\right],\\
\displaystyle \nonumber&&\mathcal{D}^{\mathcal{Y}}\left[y||y^{\prime}\right]=R\tilde{T}\sum_{i}e^{-\nu_{i}^{o}\left(\tilde{T}\right)/R\tilde{T}}\\
&&\times\left[\left\{e^{y_{i}/R\tilde{T}}-e^{y_{i}^{\prime}/R\tilde{T}}\right\}-\frac{e^{y_{i}^{\prime}/R\tilde{T}}}{R\tilde{T}}\left\{y_{i}-y_{i}^{\prime}\right\}\right],\label{BDIGY}
\end{eqnarray}
where we use Eqs. (\ref{IGTPg}), (\ref{IGmapXtoY}), (\ref{IGDTPg}) and (\ref{IGmapYtoX}). 
Note that, for the ideal gas cases, the Bregman divergences on $\mathcal{X}$ reduces to the generalized Kullback-Leibler divergence \cite{m1,05,07}. 
Therefore, if the consistency condition, Eq. (\ref{esposi}), holds, that is, if the equilibrium state exists, we can evaluate the total entropy production by the generalized Kullback-Leibler divergence through Eq. (\ref{III1}). 
This result was reported in previous work \cite{07,a3,a4,m1} based on the mass action kinetics and the local detailed balance condition. 
Also, by using the Bregman divergences on $\mathcal{Y}$, Eq. (\ref{BDIGY}), we can rephrase Prescription 1 to identify the equilibrium state for the ideal gas cases as 
\begin{remark} {\bf (}{\it {\bf Ideal} {\bf Gas}}{\bf )} 
$\left(\bm{0}\right)$ Confirm the consistency condition, Eq. (\ref{esposi}), from given stoichiometric matrices, $S$ and $O$, and the reservoir condition $(\tilde{T},\tilde{\mu})$.
$\left(\bm{1}\right)$ The convex functions $\varphi\left(x\right)$ and $\varphi^{*}\left(y\right)$ are given as Eqs. (\ref{IGTPg}) and (\ref{IGDTPg}) . 
$\left(\bm{2}\right)$ Determine the equilibrium manifold $\mathcal{V}_{\mathrm{E}\mathrm{Q}}^{\mathcal{Y}}(\tilde{\mu})$ in $\mathcal{Y}$ by solving the simultaneous equations $y_{i}S_{r}^{i}+\tilde{\mu}_{m}O_{r}^{m}=0$.
$\left(\bm{3}\right)$ Calculate the initial chemical potential $y^{0}$ by substituting a given initial density $x_{0}$ into Eq. (\ref{IGmapXtoY}). 
$\left(\bm{4}\right)$ Obtain the equilibrium state $y^{\mathrm{E}\mathrm{Q}}$ by employing the variational form, Eq. (\ref{main1}), with the Bregman divergence, Eq. (\ref{BDIGY}).
$\left(\bm{5}\right)$ The substitution of $y^{\mathrm{E}\mathrm{Q}}$ into Eq. (\ref{IGmapYtoX}) leads to the equilibrium density $x_{\mathrm{E}\mathrm{Q}}$. 
Also, the heat dissipation during a relaxation is computed by Eq. (\ref{heaty}). 
\end{remark}

\subsection{B. Mass action kinetics}
In this subsection, by using mass action kinetics as a specific form of the reaction flux density $j\left(t\right)$, we discuss, in terms of the kinetics, the chemical reaction systems composed of ideal gas.
As a result, we obtain the local detailed balance condition and find that the entropy production can be represented by the flux density.

For modeling reaction flux densities in ideal gas chemical reaction systems, we here employ mass action kinetics \cite{a1,02,03}, which is defined as follows: 
Consider a set of chemical equations, the $r$th reaction of which is represented as
\begin{equation}
\left(S_{+}\right)_{r}^{i}\mathfrak{X}_{i}+\left(O_{+}\right)_{r}^{m}\mathfrak{N}_{m}\rightleftarrows\left(S_{-}\right)_{r}^{i}\mathfrak{X}_{i}+\left(O_{-}\right)_{r}^{m}\mathfrak{N}_{m},\label{CheEq}
\end{equation}
where $\mathfrak{X}=\left\{\mathfrak{X}_{i}\right\}$ and $\mathfrak{N}=\left\{\mathfrak{N}_{m}\right\}$ are the labels of the confined and the open chemicals, respectively; also, $\left(S_{+}\right)_{r}^{i}$ and $\left(O_{+}\right)_{r}^{m}$ denote stoichiometric coefficients of the reactants in the $r$th reaction, 
whereas $\left(S_{-}\right)_{r}^{i}$ and $\left(O_{-}\right)_{r}^{m}$ are ones of the products. 
By using these coefficients, the stoichiometric matrices are represented as 
\begin{equation}
S_{r}^{i}=\left(S_{-}\right)_{r}^{i}-\left(S_{+}\right)_{r}^{i},\mbox{  } O_{r}^{m}=\left(O_{-}\right)_{r}^{m}-\left(O_{+}\right)_{r}^{m}.\label{SOmat}
\end{equation}
The law of mass action imposes the functional form of the reaction flux density of the $r$th reaction, Eq. (\ref{CheEq}), to be
\begin{eqnarray}
\nonumber j^{r}\left(x,n\right)&=&j_{+}^{r}\left(x,n\right)-j_{-}^{r}\left(x,n\right)\\
\displaystyle \nonumber&=&w_{+}^{r}\displaystyle \prod_{i,m}\left(x^{i}\right)^{\left(S_{+}\right)_{r}^{i}}\left(n^{m}\right)^{\left(O_{+}\right)_{r}^{m}}\\
&&-w_{-}^{r}\displaystyle \prod_{i,m}\left(x^{i}\right)^{\left(S_{-}\right)_{r}^{i}}\left(n^{m}\right)^{\left(O_{-}\right)_{r}^{m}}.\label{MAJ}
\end{eqnarray}
Here, $j_{\pm}^{r}\left(x,n\right)$ are the one-way fluxes which reflect the following microscopic description. 
$j_{+}^{r}\left(x,n\right)$ represents the expectation that the reaction occurs from the left to the right in Eq. (\ref{CheEq}), whereas $j_{-}^{r}\left(x,n\right)$ is the expectation that the opposite reaction happens. 
The coefficients, $w_{+}^{r}$ and $w_{-}^{r}$, are called the rate constants, which imply the conditional probability that the reaction occurs, given the condition that the involved chemicals encountered. 
The remaining product parts, $\Pi_{i,m}\left(\cdots\right)$, correspond to the probabilities that the chemicals encounter in the well-mixed situation.

Employing Eq. (\ref{MAJ}), we define the following quantity: 
\begin{equation}
\displaystyle \log\frac{j_{+}^{r}\left(x,n\right)}{j_{-}^{r}\left(x,n\right)}=\log\frac{w_{+}^{r}}{w_{-}^{r}}-\sum_{i}S_{r}^{i}\log x^{i}-\sum_{m}O_{r}^{m}\log n^{m}.\label{VI1}
\end{equation}
If the consistency condition, Eq. (\ref{esposi}), holds, the system must have equilibrium states. 
Since the flux density should vanish at equilibrium states, we get $j^{r}\left(x,n\right)=0$ ($\Leftrightarrow j_{+}^{r}\left(x,n\right)=j_{-}^{r}\left(x,n\right)$), for $x\in \mathcal{V}_{\mathrm{E}\mathrm{Q}}^{\mathcal{X}}(\tilde{T},\tilde{\mu})$ and $n$ is given by Eq. (\ref{nqeqIG}). 
By employing Eqs. (\ref{nqeqIG}) and (\ref{toricTD}), we can rewrite Eq. (\ref{VI1}) at an equilibrium state as 
\begin{eqnarray}
\displaystyle \nonumber 0&=&\displaystyle \log\frac{w_{+}^{r}}{w_{-}^{r}}-\frac{1}{R\tilde{T}}\left\{y_{i}^{P}-\nu_{i}^{0}\left(\tilde{T}\right)\right\}S_{r}^{i}\\
&&-\displaystyle \frac{1}{R\tilde{T}}\left\{\tilde{\mu}_{m}-\mu_{m}^{o}\left(\tilde{T}\right)\right\}O_{r}^{m},\label{VI2}
\end{eqnarray}
where $y^{P}\in \mathcal{V}_{\mathrm{E}\mathrm{Q}}^{\mathcal{Y}}(\tilde{\mu})$ and we use $US=0$. 
Since $y^{P}$ is a particular solution of the simultaneous equations $y_{i}^{P}S_{r}^{i}+\tilde{\mu}_{m}O_{r}^{m}=0$, we obtain, from Eq. (\ref{VI2}), 
\begin{equation}
\displaystyle \log\frac{w_{+}^{r}}{w_{-}^{r}}=-\frac{1}{R\tilde{T}}\left\{\nu_{i}^{0}\left(\tilde{T}\right)S_{r}^{i}+\mu_{m}^{o}\left(\tilde{T}\right)O_{r}^{m}\right\},\label{LDB}
\end{equation}
which is known as the local detailed balance condition \cite{06,07}. 
This condition bridges kinetics and thermodynamics. 
%Furthermore, we should note that this condition holds, irrespective of the consistency condition Eq. (\ref{esposi}).

In the ideal gas case, the densities of open chemicals are constant (see Eq. (\ref{nqeqIG})). 
For notational simplicity, effective rate constants are often employed \cite{03}, which are defined as
\begin{eqnarray}
&&\displaystyle \hat{w}_{+}^{r}:=w_{+}^{r}\prod_{m}\left(n^{m}\right)^{\left(O_{+}\right)_{r}^{m}},\\
&&\displaystyle \hat{w}_{-}^{r}:=w_{-}^{r}\prod_{m}\left(n^{m}\right)^{\left(O_{-}\right)_{r}^{m}}.
\end{eqnarray}
In this case, the chemical equation, Eq. (\ref{CheEq}), reduces to the effective one: 
\begin{equation}
\left(S_{+}\right)_{r}^{i}\mathfrak{X}_{i}\rightleftarrows\left(S_{-}\right)_{r}^{i}\mathfrak{X}_{i},
\end{equation}
and the local detailed balance condition, Eq. (\ref{LDB}), can be read as 
\begin{equation}
\displaystyle \log\frac{\hat{w}_{+}^{r}}{\hat{w}_{-}^{r}}=-\frac{1}{R\tilde{T}}\left\{\nu_{i}^{0}\left(\tilde{T}\right)S_{r}^{i}+\tilde{\mu}_{m}O_{r}^{m}\right\},
\end{equation}
where we use Eq. (\ref{nqeqIG}). 

Finally, we confirm that the local detailed balance condition guarantees that the system satisfies the second law. 
The differentiation of the total entropy, Eq. (\ref{enttotXi}), with respect to time $t$, leads to
\begin{eqnarray}
\displaystyle \nonumber\frac{d}{dt}\Sigma^{\mathrm{t}\mathrm{o}\mathrm{t}}\left(\xi\left(t\right)\right)&=&-\displaystyle \frac{\Omega}{\tilde{T}}\{\nu_{i}^{0}(\tilde{T})S_{r}^{i}+R\tilde{T}\sum_{i}S_{r}^{i}\log x^{i}\left(t\right)\\
&&+\tilde{\mu}_{m}O_{r}^{m}\}j^{r}\left(t\right),\label{VI3}
\end{eqnarray}
where we use Eq. (\ref{IGmapXtoY}). 
By substituting the local detailed balance condition, Eq. (\ref{LDB}), into Eq. (\ref{VI3}), we get 
\begin{eqnarray}
\displaystyle \nonumber\frac{d\Sigma^{\mathrm{t}\mathrm{o}\mathrm{t}}}{dt}&=&\displaystyle \Omega R\left\{\log\frac{w_{+}^{r}}{w_{-}^{r}}-\sum_{m}O_{r}^{m}\log n^{m}\right.\\
&&\left.-\displaystyle \sum_{i}S_{r}^{i}\log x^{i}\left(t\right)\right\}j^{r}\left(t\right),
\end{eqnarray}
where we use Eq. (\ref{nqeqIG}). 
From Eq. (\ref{VI1}), we obtain 
\begin{equation}
\displaystyle \frac{d\Sigma^{\mathrm{t}\mathrm{o}\mathrm{t}}}{dt}=\Omega R\sum_{r}\left\{j_{+}^{r}\left(t\right)-j_{-}^{r}\left(t\right)\right\}\log\frac{j_{+}^{r}\left(t\right)}{j_{-}^{r}\left(t\right)}\geq 0,\label{totj}
\end{equation}
which guarantees non-negativity of the entropy production rate. 
Also, since the equality holds if and only if $j_{+}=j_{-}$ (that is $j=0$), the entropy production rate is strictly positive except for equilibrium states. 
That is precisely the second law.
In addition, the representation by the flux densities, Eq. (\ref{totj}), is often employed to evaluate the entropy production in the chemical reaction systems \cite{02,03,04,05,06,07,08,09}.

\section{VI. Summary and discussion}

We have established the Hessian geometric structure in chemical thermodynamics of CRNs. 
We have derived the existence and uniqueness condition of the equilibrium state, which is determined by the intersection of equilibrium and stoichiometric manifolds. 
Also, the entropy production during a relaxation to the equilibrium state is evaluated by the Bregman divergence. 
Furthermore, the equilibrium state is characterized by four distinct minimization 
problems of the divergence, two of which are in the density space and the other two are in the chemical potential space. 
For the ideal gas cases, we have confirmed that our Theorem 2 reduces to Birch's theorem, 
and the entropy production represented by the divergence coincides with the generalized Kullback-Leibler divergence; 
the additional assumption of the mass action kinetics leads to the local detailed balance condition.

Although we have only treated the isochoric ideal gas cases in Sec. V, the application is straightforward to conventional CRNs appearing in isobaric ideal-dilute-solution situations in chemistry and biology. 
To move from an isochoric to an isobaric situation, we replace the Helmholtz free energy with the Gibbs one. 
At this step, one may be concerned that the volume of the system can change under a constant pressure. 
However, we can effectively identify the isobaric situation with the isochoric one, because the solvent dominates the volume and the amount of solvent is constant in the reaction dynamics. 
Thus, the Gibbs free energy is obtained just by modifying the standard chemical potentials in Eq. (\ref{IDHF}) \cite{01,07}. 
This direct correspondence between Helmholtz and Gibbs free energies originates from the fact that we can regard the solvent as the background of the reaction dynamics.  

However, there exist situations, such as cellular growth, which do not have the simple correspondence between isochoric and isobaric free energies.
In this situation, the volume $\Omega$ is no longer constant with time, and thus the system may not have any conserved quantities. 
Due to that, the homogeneity of the entropy function (see Eq. (\ref{fe})) 
gives a non-trivial impact to the structure of our theory, and a further extension is required \cite{f1}. 

Much work has been devoted to interpret thermodynamics with geometric frameworks \cite{g4,g5,g6,aa1,aa2}. 
They revealed the geometric dual structure by Legendre transformations in thermodynamics.
However, if non-trivial constraints such as stoichiometric ones enter the problem, 
the constraints introduce important submanifolds (equilibrium and stoichiometric manifolds) into the Legendre dual spaces.
We have clarified how the resulting Hessian geometric structure enables us to handle the complex constraints in CRNs. 
We have also demonstrated that the characteristic thermodynamic properties obtained for mass action systems with the local detailed balance condition emerge from this fundamental structure without assuming any of them.  
%The contribution of our theory resides in that we can handle multi-dimensional reservoir conditions generating complex constraints in CRNs. 
%These complex constraints introduce the submanifolds into the Legendre dual structure.   
Not limited to CRNs, such geometric structure with the submanifolds can appear in a wide variety of systems with complex constraints, which implies general applicability of our theory. 

In this work, we have only dealt with the cases that the system converges to the equilibrium state, that is, the reservoir satisfies the condition, Eq. (\ref{esposi}).
Otherwise, the equilibrium state does not exist, and the total entropy keeps increasing and finally diverging. 
Even in such cases, it is known that the system may converge to a certain stable state in a time evolution, which is called the nonequilibrium steady state (NESS) \cite{05,06,07,08,09}. 
A typical example of NESS in CRNs is the complex-balanced state \cite{m1,m4,05,07}.  
However, we can not characterize NESS solely by the entropy function, because the variational form based on the entropy maximization as in Eq. (\ref{VPXi}) can no longer be employed. 
The extension of our geometric structure to the cases of NESS is future work \cite{Kobaflux}.  

\section{Acknowledgement}
This research is supported by JSPS KAKENHI Grant Numbers 19H05799 and 21K21308, and by JST CREST JPMJCR2011 and JPMJCR1927.

%$=-\displaystyle \frac{\Omega}{\tilde{T}}\left\{-R\tilde{T}\log\frac{w_{+}^{r}}{w_{-}^{r}}-\mu_{m}^{o}\left(\tilde{T}\right)O_{r}^{m}+\tilde{\mu}_{m}O_{r}^{m}+R\tilde{T}\sum_{i}S_{r}^{i}\log x^{i}\right\}j^{r}\left(t\right)=-\Omega R\left\{-\log\frac{w_{+}^{r}}{w_{-}^{r}}+\sum_{m}O_{r}^{m}\log n^{m}+\sum_{i}S_{r}^{i}\log x^{i}\right\}j^{r}\left(t\right)=\Omega R\sum_{r}j^{r}\left(t\right)\log\frac{j_{+}^{r}}{j_{-}^{r}}>0$

\appendix
\section{Appendix A}
Here, we derive $\partial\varphi(\tilde{T},\tilde{\mu};x)/\partial\tilde{T}=-\sigma_{\mathrm{Q}\mathrm{E}\mathrm{Q}}\left(x\right)$. 
By substituting the definitions of the thermodynamic potentials, Eqs. (\ref{TPm}) and (\ref{TPf}), into Eq. (\ref{TPg}), we get 
\begin{eqnarray}
\displaystyle \nonumber\varphi\left[\tilde{T},\tilde{\mu};x\right]&=&-\displaystyle \max_{\epsilon,n}\left\{\tilde{T}\sigma\left[\epsilon,n,x\right]-\epsilon+\tilde{\mu}_{m}n^{m}\right\}\\
\nonumber&=&-\tilde{T}\sigma\left[\epsilon_{\mathrm{Q}\mathrm{E}\mathrm{Q}},n_{\mathrm{Q}\mathrm{E}\mathrm{Q}},x\right]+\epsilon_{\mathrm{Q}\mathrm{E}\mathrm{Q}}\left(\tilde{T},\tilde{\mu};x\right)\\
&&-\tilde{\mu}_{m}n_{\mathrm{Q}\mathrm{E}\mathrm{Q}}^{m}\left(\tilde{T},\tilde{\mu};x\right),\label{A1}
\end{eqnarray}
where we use Eq. (\ref{VPEN}). 
Therefore, we obtain $\partial\varphi(\tilde{T},\tilde{\mu};x)/\partial\tilde{T}=-\sigma\left[\epsilon_{\mathrm{Q}\mathrm{E}\mathrm{Q}},n_{\mathrm{Q}\mathrm{E}\mathrm{Q}},x\right]=-\sigma_{\mathrm{Q}\mathrm{E}\mathrm{Q}}\left(x\right),$ where we note that the implicit differentiations with respect to $\epsilon_{\mathrm{Q}\mathrm{E}\mathrm{Q}}$ and $n_{\mathrm{Q}\mathrm{E}\mathrm{Q}}$ vanish, due to the critical equation of Eq. (\ref{A1}). 
In addition, from the same reason, we get $n_{\mathrm{Q}\mathrm{E}\mathrm{Q}}(\tilde{T},\tilde{\mu};x)$ and $\epsilon_{\mathrm{Q}\mathrm{E}\mathrm{Q}}(\tilde{T},\tilde{\mu};x)$ as 
\begin{eqnarray}
n_{\mathrm{Q}\mathrm{E}\mathrm{Q}}^{m}\displaystyle \left(\tilde{T},\tilde{\mu};x\right)&=&-\displaystyle \frac{\partial\varphi\left(\tilde{T},\tilde{\mu};x\right)}{\partial\tilde{\mu}_{m}},\label{nQEQ}\\
\displaystyle \epsilon_{\mathrm{Q}\mathrm{E}\mathrm{Q}}\left(\tilde{T},\tilde{\mu};x\right)&=&\displaystyle \varphi\left[\tilde{T},\tilde{\mu};x\right]-\tilde{\mu}_{m}\frac{\partial\varphi}{\partial\tilde{\mu}_{m}}-\tilde{T}\frac{\partial\varphi}{\partial\tilde{T}}.
\end{eqnarray}

\section{Appendix B}
The differentiations of the free-energy density $f[\tilde{T};n,x]$ with respect to $n$ and $x$ lead to the forms of chemical potentials as 
\begin{eqnarray}
\mu_{m}\left(\tilde{T};n,x\right)&=&\mu_{m}^{o}\left(\tilde{T}\right)+R\tilde{T}\log n^{m},\label{AppB4}\\
y_{i}\left(\tilde{T};n,x\right)&=&\nu_{i}^{o}\left(\tilde{T}\right)+R\tilde{T}\log x^{i}.\label{AppB5}
\end{eqnarray}
From these forms, we can find that $\mu_{m}^{o}(\tilde{T})$ and $\nu_{i}^{o}(\tilde{T})$ are the chemical potentials standardized at $n^{m}=1$ and $x^{i}=1$ in the chosen physical units for all $m$ and $i$, that is, $\mu_{m}^{o}(\tilde{T}):=\mu_{m}(\tilde{T};1,1)$ and $\nu_{i}^{o}(\tilde{T}):=y_{i}(\tilde{T};1,1)$. 
If one wants to standardize at arbitrary concentrations, $\bar{n}^i$ and $\bar{x}^i$, the forms of chemical potentials can be written as 
\begin{eqnarray}
\mu_{m}\left(\tilde{T};n,x\right)&=&\bar{\mu}_{m}^{o}\left(\tilde{T}\right)+R\tilde{T}\log \frac{ n^{m}}{\bar{n}^m},\label{AppB1}\\
y_{i}\left(\tilde{T};n,x\right)&=&\bar{\nu}_{i}^{o}\left(\tilde{T}\right)+R\tilde{T}\log \frac{x^{i}}{\bar{x}^{i}}.\label{AppB2}
\end{eqnarray}
Here, the standard chemical potentials are modified as 
\begin{eqnarray}
\bar{\mu}_{m}^{o}(\tilde{T})&=&\mu_{m}(\tilde{T};\bar{n},\bar{x})=\mu_{m}^{o}\left(\tilde{T}\right)+R\tilde{T}\log \bar{n}^{m},\label{AppB6}\\
\bar{\nu}_{i}^{o}(\tilde{T})&=&y_{i}(\tilde{T};\bar{n},\bar{x})=\nu_{i}^{o}\left(\tilde{T}\right)+R\tilde{T}\log \bar{x}^{i}.\label{AppB7}
\end{eqnarray}
Furthermore, the functional form of free energy in Eq. (\ref{IDHF}) should be represented as
\begin{eqnarray}
\displaystyle \nonumber f\left[\tilde{T};n,x\right]&=&n^{m}\displaystyle \bar{\mu}_{m}^{o}\left(\tilde{T}\right)+R\tilde{T}\sum_{m}\left\{n^{m}\log \frac{n^{m}}{\bar{n}^m}-n^{m}\right\}\\
&&+x^{i}\displaystyle \bar{\nu}_{i}^{o}\left(\tilde{T}\right)+R\tilde{T}\sum_{i}\left\{x^{i}\log \frac{x^{i}}{\bar{x}^{i}}-x^{i}\right\}.\label{AppB3}
\end{eqnarray}
Taking the physical dimensionality into account, the representations, Eqs. (\ref{AppB1}), (\ref{AppB2}) and (\ref{AppB3}), are more suitable, because the insides of logarithms become dimensionless. 
However, we use the representations, Eqs. (\ref{IDHF}), (\ref{AppB4}) and (\ref{AppB5}), in the main text, for notational simplicity. 

In chemistry, the chemical potential is often standardized by pressure. 
It is straightforward to switch our standard chemical potentials, $\mu_{m}^{o}$ and $\nu_{i}^{o}$, to the common ones, $\hat{\mu}_{m}^{o}$ and $\hat{\nu}_{i}^{o}$, as follows. 
For the standard partial pressures, $\bar{\Pi}^m$ and $\bar{P}^i$, the equations of state can be represented as $\bar{n}^m=\bar{\Pi}^m/R\tilde{T}$ and $\bar{x}^i=\bar{P}^i/R\tilde{T}$. 
By substituting them into Eqs. (\ref{AppB6}) and (\ref{AppB7}), we obtain  
\begin{eqnarray}
\displaystyle \hat{\mu}_{m}^{o}\left(\tilde{T}\right)&=&\displaystyle \mu_{m}^{o}\left(\tilde{T}\right)+R\tilde{T}\log\frac{\bar{\Pi}^m}{R\tilde{T}},\\
\displaystyle \hat{\nu}_{i}^{o}\left(\tilde{T}\right)&=&\displaystyle \nu_{i}^{o}\left(\tilde{T}\right)+R\tilde{T}\log\frac{\bar{P}^i}{R\tilde{T}}.
\end{eqnarray}
Then, the chemical potentials, Eqs. (\ref{AppB1}) and (\ref{AppB2}), can be rewritten as 
\begin{eqnarray}
\displaystyle \mu_{m}\left(\tilde{T};\Pi,P\right)&=&\hat{\mu}_{m}^{o}\left(\tilde{T}\right)+R\tilde{T}\log\frac{{\Pi}^m}{\bar{\Pi}^m},\\
\displaystyle y_{i}\left(\tilde{T};\Pi,P\right)&=&\hat{\nu}_{i}^{o}\left(\tilde{T}\right)+R\tilde{T}\log\frac{P^i}{\bar{P}^i},
\end{eqnarray}
where ${\Pi}^m$ and $P^i$ are the partial pressures for $n^m$ and $x^i$, respectively.  



\begin{thebibliography}{00}
\bibitem{n0}  The isochoric cases are theoretically identical to conventional CRNs in isobaric ideal-dilute-solution cases. For details, see Summary and Discussion in Sec. VI.
\bibitem{n1}  To be more precise, $\Sigma^{\mathrm{t}\mathrm{o}\mathrm{t}}\left(\tau\right)$ must be a strictly increasing function except for its maximum.
\bibitem{n2}  The critical equations of the variational form, Eq. (\ref{VPEN}), are $\partial\sigma/\partial\epsilon-\tilde{T}=0$ and $\partial\sigma/\partial n^{m}-\tilde{\mu}^{m}=0$ for all $m$. 
%In the ordinary situation, the ranges of $\partial\sigma/\partial\epsilon$ and $\partial\sigma/\partial n^{m}$ are $\mathbb{R}_{>0}$ and $\mathbb{R}$ respectively; the maximum of Eq. (\ref{VPEN}) exists for $\tilde{T}>0, \tilde{\mu}\in \mathbb{R}$. 
In the ordinary situation, we have 
\begin{eqnarray*}
&&\left\{ \left( \partial \sigma(\epsilon, n, x)/\partial\epsilon, \partial \sigma(\epsilon, n, x)/\partial n \right) \right.\\&&\left.|(\epsilon, n, x) \in \mathbb{R} \times \mathbb{R}^{\mathcal{N}_N}_{>0} \times \mathbb{R}^{\mathcal{N}_X}_{>0} \right\} = \mathbb{R}_{>0} \times \mathbb{R}^{\mathcal{N}_{N}}, 
\end{eqnarray*}
which is mathematically an assumption of thermodynamics. Therefore, the maximum of Eq. (\ref{VPEN}) exists for $\tilde{T}>0, \tilde{\mu}\in \mathbb{R}$.
\bibitem{n3}  The inverse transformation of this type Legendre transformation is given by $f[\tilde{T};n,x]=\mathrm{max}_{\tilde{\mu}}\{\varphi[\tilde{T},\tilde{\mu};x]+n^{m}\tilde{\mu}_{m}\}$.
\bibitem{n5}  The common definition of the affinity (the thermodynamic force) $\hat{A}$ may be the differentiation of the total entropy $\Sigma^{\mathrm{t}\mathrm{o}\mathrm{t}}\left(\xi\right)$ in Eq. (\ref{enttotXi}) with respect to the extent of reaction $\xi$, that is, $\hat{A}=(\Omega/\tilde{T})A$. 
However, since the volume $\Omega$ and the temperature $\tilde{T}$ are constant in our setup, we refer to $A$ as the affinity.
\bibitem{n6}  The differentiation of the function with respect to $x$ leads to $y^P_i-\partial_i\varphi(x)$. Some $x$ exists such that $y^P_i-\partial_i\varphi(x)=0$, because the finite chemical potential $y^P$ has the finite corresponding density $x_P=\partial\varphi^*(y^P)$ in the ordinary thermodynamics. Thus, the function is bounded above on $\mathcal{X}$.
\bibitem{n4}  The uniqueness here is different from one in the fast dynamics. For the fast dynamics, the quasi-equilibrium state is uniquely determined, irrespective of an initial condition.
\bibitem{na} If we select another particular solution $y'^{P}$ to the simultaneous equations $y_i S^i_r = -\tilde{\mu}_m O^m_r$, we can write it as $y'^{P}=y^{P}+u$ where $u \in {\rm Ker}[S^T]$. By taking $x(t)-x_0\in {\rm Im}[S]$ into account, we find that the term, $y'^{P}_i (x^i(t)-x^i_0)$, in Eq. (\ref{prepent}) is equivalent to $y_i^P (x^i(t)-x^i_0)$, because $u_i(x^i(t)-x^i_0) = 0$.
\bibitem{n7}  We should note that this orthogonality is defined by the inner product with the Hessian metric $g_{ij}:=\partial^2 \varphi/\partial x^i \partial x^j$.
\end{thebibliography}
\end{document}